\documentclass[aps,eqsecnum,amsmath,onecolumn,groupedaddress,superscriptaddress,notitlepage,nofootinbib]{revtex4-1}
\usepackage[utf8]{inputenc}
\usepackage{indentfirst}
\usepackage{amssymb}
\usepackage{amsmath,latexsym}
\usepackage{graphicx}
\usepackage{caption}
\usepackage{subcaption}
\usepackage{float}
\usepackage[margin=1.0in]{geometry}
\usepackage[mathscr]{euscript}
\usepackage{color}
\usepackage{xcolor}
\usepackage{soul}
\usepackage{slashed}
\graphicspath{{images/}}
\usepackage[mathscr]{euscript}
\usepackage{subcaption}
\usepackage{color}
\DeclareMathOperator{\Tr}{Tr}

\newcommand{\beq}{\begin{equation}}
\newcommand{\eeq}{\end{equation}}
\newcommand{\bea}{\begin{eqnarray}}
\newcommand{\eea}{\end{eqnarray}}
\newcommand{\dbar}{\mathchar'26\mkern-9mu d}
\newcommand{\vecq}{\mathbf q}

\newcommand{\vecr}{\mathbf r}
\newcommand{\vecS}{\mathbf S}

\newcommand{\vth}{v_{\rm th}}

\def\beqs#1\eeqs{\beq\begin{split} #1 \end{split}\eeq}

\def\Tr {\mathop{\hbox{Tr}}}

\usepackage[colorlinks=true,backref=false, linktocpage=true,
citecolor=blue,urlcolor=blue,linkcolor=blue,pdfpagemode=UseOutlines]{hyperref}

\hypersetup{%
  bookmarksnumbered=true,
  pdftitle = {},
  pdfsubject = {},
  pdfauthor = {},
  pdfkeywords = {}
}

\usepackage{microtype}

\newcommand{\nn}{\nonumber}
\newcommand{\eq}[1]{Eq.~(\ref{#1})}
\newcommand{\fig}[1]{Fig.~\ref{#1}}

\begin{document}
\title{Neutrino-nucleon scattering in the neutrino-sphere}

\author{Paulo F. Bedaque}
\email{bedaque@umd.edu}
\affiliation{Department of Physics,
University of Maryland, College Park, MD 20742}

\author{Sanjay Reddy}
\email{sareddy@uw.edu}
\affiliation{Institute of Nuclear Theory, University of Washington, Seattle WA 98195}

\author{Srimoyee Sen}
\email{srimoyee08@gmail.com}
\affiliation{Institute of Nuclear Theory, University of Washington, Seattle WA 98195}

\author{Neill C. Warrington}
\email{ncwarrin@umd.edu}
\affiliation{Department of Physics,
University of Maryland, College Park, MD 20742}

\date{\today}

\begin{abstract}
We calculate the differential scattering rate for thermal neutrinos in a hot and dilute gas of interacting neutrons using linear response theory. The dynamical structure factors for density and spin fluctuations of the strongly interacting neutron matter, expected in the neutrino decoupling regions of supernovae and neutron star mergers, are calculated in the virial expansion for the first time. Correlations due to nucleon-nucleon interactions are taken into account using a pseudo-potential that reproduces measured nucleon-nucleon phase shifts, and we find that attractive s-wave interactions  enhance the density response and suppress the spin response of neutron matter. The net effect of neutron correlations is to strongly suppress backscattering. Moreover, we find nearly exact scaling laws for the response functions, valid for the range $T=5-10$ MeV and $q<30$ MeV, allowing us to obtain analytic results for the dynamic structure factors at second-order in the fugacity of the neutron gas. We find that the modification of scattering rates depends on the energy and momentum exchanged, implying that dynamical structure factors are essential to describe neutrino decoupling in supernovae and neutron star mergers.
\end{abstract}


\maketitle

\section{Introduction}
The energy spectrum of neutrinos emerging from supernovae and neutron star mergers influence the supernova explosion mechanism, nucleosynthesis, and their detectability in terrestrial neutrino detectors. An accurate description of neutrino interactions in hot dense matter encountered in these extreme phenomena is essential to make reliable predictions for the neutrino spectrum and luminosity and has been studied extensively (see \cite{Burrows:2004vq} for a recent review). It is known that strong interactions between nucleons and electromagnetic interactions between nucleons and charged leptons can alter neutrino scattering rates and influence the temporal and spectral features of neutrino emission from supernovae \cite{Sawyer:1975,Iwamoto:1982zp,Horowitz:1990it,Burrows:1998cg,Reddy:1998hb}. 

In this article we shall focus on neutrino interactions in matter at moderate density ($\rho\simeq 10^{11}-10^{13}~ \text{g}/{\text{cm}^3}$) and high temperature ($T=5-10$ MeV), since these are conditions encountered in the neutrino-sphere region where neutrino decouple from matter and their energy spectrum is determined.  Under these conditions nucleons form a dilute gas, 
and the fugacity of nucleons $z=e^{\mu/T}$, where $\mu$ is the nucleon chemical potential, is a useful expansion parameter. This has been exploited to calculate the equation of state (EOS) directly in terms of the measured nucleon-nucleon phase shifts using the well-known virial expansion \cite{Horowitz:2005zv,Horowitz:2006pj}. Further, since response functions in the long-wavelength limit are related to thermodynamic derivates, the virial EOS has been used to obtain neutrino scattering rates in dilute matter by neglecting corrections that depend on the energy and momentum transfer in neutrino-nucleon scattering. The main objective of this study is to assess how strong interaction corrections to the neutrino-nucleon scattering depend on the energy and momentum transfer.


\section{Neutrino scattering rate in a neutron gas}
Although matter encountered in the neutrino-sphere contains neutrons, protons, electrons, and perhaps even small traces of light nuclei, in the following we shall focus on neutrino scattering in a pure neutron gas. This will allow us to establish the formalism and examine in detail the effects due to nuclear interactions without the added complexity of multi-component systems with electrons and protons, where long-range electromagnetic interactions will also need to be accounted. Further, since matter in the neutrino-sphere is close to $\beta-$equilibrium, with negligibly small neutrino chemical potential, the fraction of charged particles (electrons, proton, and light nuclei) is typically much less than $10$\%, and neutrons dominate the scattering opacity.   

The differential scattering rate of low energy neutrinos in a non-relativistic gas of neutrons is given by 
\beq\label{eq:diff2_rate}
\frac{d\Gamma(E_\nu)}{d\cos{\theta}~dq_0} = \frac{G_F^2 }{4\pi^2} ~(E_\nu-q_0)^2~\left[
c^2_V ~(1+\cos\theta)~S_V(q_0, \vecq)+
c_A^2(3-\cos\theta)S_A(q_0, \vecq)
\right]
\eeq
where $E_\nu$ is the energy of the incoming neutrino, $q_0$ is the energy transfer to the medium, and $\theta$ is the angle between the incoming and outgoing neutrino. The momentum transfer to the medium $\vecq$ is constrained by kinematics to satisfy $|\vecq| = \sqrt{4E_{\nu}(E_{\nu}-q_0)\sin^2{(\theta/2)}+q_0^2}$. The neutral current vector and axial vector coupling constants for the neutron are $c_V=-1/2$ and $c_A=-(g_A-\Delta S)/2$, respectively, where $g_A\approx-1.27 $ and $\Delta S \approx 0$. $S_{V,A}(q_0,\vecq)$ are the density and spin structure factors defined by
\bea
S_V(q_0, \vecq) &=& \int dt d^3\vecr\ e^{iq_0t-i\vecq . \vecr}\langle \delta n(t,\vecr)  \delta n(0,0)\rangle \nn\\
S_A(q_0, \vecq) &=& \int dt d^3\vecr\ e^{iq_0t-i\vecq . \vecr}\langle \delta \vecS(t,\vecr)  \delta \vecS(0,0)\rangle
\label{eq:structure_factors}
\eea where the thermal average is $\langle . \rangle = \Tr(e^{-\beta H} . )/ \Tr e^{-\beta H} $ and $\delta n = n-\langle n
\rangle$ ($\delta \vecS = \vecS - \langle \vecS\rangle$) are the fluctuations of the density (spin). The approximations leading to \eq{eq:diff2_rate} are only that the weak interaction is treated at first-order in the coupling, and that neutrons  are non-relativistic. The latter greatly simplifies the calculation since,  to order $v^0$ where $v$ is the nucleon velocity, the nucleon vector current reduces to $\bar{\psi}\gamma_\mu\psi \rightarrow \delta_{\mu0}\psi^\dagger \psi$  and the axial current  reduces to its spatial part $\bar{\psi}\gamma_\mu \gamma_5 \psi \rightarrow  \delta_{\mu i}\psi^\dagger \sigma_i \psi$, resulting in an expression entirely determined by the fluctuations of density and spin.

While the weak interactions between neutrinos and nucleons is perturbative, the interactions among nucleons is not, especially at the temperatures and densities encountered in the neutrino-sphere. As a consequence, methods needed to calculate  the exact density and spin dynamic structure factors of a non-perturbative dense many-body system are still lacking. Perturbation theory in the strength of the strong interaction fails and non-perturbative many-body computational methods such as Quantum Monte Carlo (QMC), which have been useful to obtain ground state energies and thermodynamic properties of strongly interacting dense Fermi systems, cannot be directly used to calculate the frequency dependence of response functions because they are formulated in imaginary time. Further, interactions between nucleons at short-distances is poorly known, and three and higher-body forces begin to play a role at and above nuclear saturation density ($\rho_{\rm sat} \approx 2.5 \times 10^{14}$ g/cm$^3$).  

Before we calculate the dynamic structure functions and discuss the approximations involved, we present results that can be obtained with only static information about the density and spin correlation functions. Integrating over kinematically allowed energy transfers we can rewrite \eq{eq:diff2_rate} as 
\begin{align}\label{eq:diff_rate}
\frac{d\Gamma(E_\nu)}{dq} = \frac{G_F^2~q}{2\pi^2} & \left[
 c^2_V \tilde{S}_V(q)\left( 1- \frac{q^2}{4E^2_\nu} -\frac{\omega_V}{E_\nu}  + \frac{\omega^2_V}{4E^2_\nu} \right) +  c_A^2\tilde{S}_A(q)\left(1+  \frac{q^2}{4E^2_\nu}-\frac{\omega_A}{E_\nu}- \frac{\omega^2_A}{4E^2_\nu}\right) \right]\,,
\end{align}
where 
\begin{align}
\tilde{S}_{V/A}(q)&=\int^{{\rm min}[2E_\nu-q,q]}_{-q}~dq_0 ~S_{V/A}(q_0, q)\,, \\
\omega^n_{V/A} &= \frac{1}{\tilde{S}_{V/A}(q)}\int^{{\rm min}[2E_\nu-q,q]}_{-q}~dq_0~q^n_0 ~S_{V/A}(q_0, q)\,.
\label{eq:omega}
\end{align}
and $q=|\vecq|$ is the magnitude of the momentum transfer. The functions $\tilde{S}_{V/A}(q)$ are closely related to the static structure functions  
\beq
S_{V/A}(q)=\int^{\infty}_{-\infty}~dq_0 ~S_{V/A}(q_0, q)\,,
\eeq
and $\tilde{S}_{V/A}(q) \simeq S_{V/A}(q) $ only if a significant fraction of the response resides in the region where $ -q< q_0 < {\rm min} [2E_\nu-q,q]$.  For non-relativistic and  non-interacting nucleons, the characteristic energy transfer is of order $|q_0| \simeq \vth ~q$, where $\vth \simeq \sqrt{T/M}$ is the thermal velocity of non-degenerate nucleons with mass $M \gg T$. In the temperature range we are interested in ($T \simeq 5-10$ MeV), the thermal velocity is indeed small and $\tilde{S}_{V/A}(q) \approx S_{V/A}(q) $ should be a good approximation. However, interactions can alter this, allowing the response to peak at larger values of $|q_0|$, and in general  $\tilde{S}(\vecq) <S(\vecq) $, implying that some dynamical information  is needed to obtain a quantitative description of the scattering rates.  

The integral in \eq{eq:omega} that defines $\omega_{V/A}$  is closely related to the f-sum rule \cite{Mahan} which states that 
\beq
\int_{-\infty}^{\infty}\frac{dq_0}{2\pi}q_0 S_{\cal O}(q_0,q) =\langle [[{\cal H},{\cal O}],{\cal O}]\rangle \,,
\label{eq:FSumGeneral}
\eeq
where ${\cal O}=\psi^\dagger \psi$ (for the density response) or  ${\cal O}=\psi^\dagger \sigma_i \psi$ (for the spin response), and ${\cal H}$ is the nuclear Hamiltonian. 
When a large fraction of the response is kinematically accessible, the f-sum rule for the density response requires that $\omega_{V} =q^2/2M$,  even in the presence of interactions, as a computation of the double commutator shows. Hence we  expect $\omega_{V} \ll  E_\nu$ since typical $q\simeq E_\nu \ll M$.  However, we note that since spin is not conserved by nuclear interactions, the f-sum rule for the spin response does not vanish in the long-wavelength limit \cite{Olsson:2002yu}. One cannot guarantee that  $\omega_{A} \ll  E_\nu$ even for non-relativistic nucleons, and calculations of the dynamical response including components of ${\cal H}$ that do not commute with the spin operator are needed to determine $\omega_{A} $ \cite{Raffelt:1993ix,Lykasov:2008yz,Shen:2012sa}. 

Nonetheless, it is common practice to adopt the elastic approximation and, in the limit $(\omega_{V/A}/E_\nu\rightarrow 0)$ one obtains a simpler formula  for the differential scattering rate :
\beq \label{eq:diff_elas}
\frac{d\Gamma(E_\nu)}{d\cos{\theta}} = \frac{G_F^2 }{4\pi^2} ~E_\nu^2~\left[
c^2_V ~(1+\cos\theta)~S_V(\vecq)+ c_A^2(3-\cos\theta)S_A(\vecq) \right]\,,
\eeq
which is widely used in the literature to describe neutrino-nucleon scattering at low energy \cite{Horowitz:2006pj}.  Another approximation that greatly simplifies calculations is to also neglect the momentum transfer and replace $S_{V/A}(q)$ by $S_{V/A}(0)$. Since the latter is a long-wavelength property it can be related to the equation of state \cite{Horowitz:2006pj}.  The neglect of the momentum dependence is justified when the momentum transfer is small compared to the typical thermal nucleon momentum $p_{\rm thm} \simeq \sqrt{6 M T}$. For strongly correlated nucleons other smaller momentum scales associated with correlations between particles arise and it is {\it a priori} unclear whether the replacement $S_{V/A}(q)$ by $S_{V/A}(0)$ is a good approximation. For these reasons, and to obtain a quantitative description of how corrections to neutrino scattering due to correlations depend on energy and momentum transfer we calculate the dynamical structure function.    
\section{Method}
We will now discuss the calculation of the dynamical structure factors and the approximations involved.  As noted earlier, the relatively low density and high temperature encountered in the neutrino-sphere provides a useful small expansion parameter: the fugacity of the gas defined as $z=e^{\beta\mu}$ where $\mu$ is the chemical potential, and $\beta=1/T$ is the inverse temperature. When $z \ll 1 $ thermodynamic and linear response properties of gases can be obtained in the the virial expansion where observables are expressed as a power series in $z$. Since the fugacity is proportional to the number density at lowest order:
\beq
n \simeq  2\left( \frac{MT}{2\pi} \right)^{3/2} z,
\eeq 
the condition $z\alt 1/10$ implies that $n \alt  0.0005~(T/5\ {\rm MeV})^{3/2}$ fm$^{-3}$ or $\rho  \alt (T/5\ {\rm MeV})^{3/2} 10^{12}$ g/cm$^3$.

The way we treat the strong interactions involves an uncontrolled but well motivated approximation.  Particle-hole loops are suppressed by powers of the fugacity $z$ but particle-article loops are not \cite{Bedaque:2002xy}. Since the nuclear interactions are not perturbatively small, particle-particle loops need to be resummed to all orders. The calculation of all diagrams involving up to two particle-hole loops and an arbitrary number of particle-particle loops is very involved. However, if we drop all the particle-particle loops and, at the same time, substitute the interaction to have a pseudo-potential vertex of the form \cite{Rrapaj:2014yba}:
\beq
V(p,p') = \frac{4\pi}{M} \left (\frac{\delta(p)}{p} + \frac{\delta(p')}{p'}  \right),
\eeq where $\delta(p)$ is the phase shift, and $p$ and $p'$ are the incoming and outgoing relative momenta, one reproduces, up to  order $z^2$, the correct thermodynamics quantities as given by the Beth-Uhlenbeck formula. Thus, for simplicity,  we describe the neutron-neutron interactions by the pseudo-potential remembering to drop the particle-particle loops. This approach has the feature of including the correct, experimentally determined phase shifts as opposed to an approximation to it. On the other hand it is not rigorous in the sense that it is possible that the pseudo-potential, despite giving exact results for static quantities, does not reproduce the exact value for non-static ones.

In addition to the approximations described above, important for our ability to compute, we will make the following approximations only to keep the calculations simple.  First, we do not include  the higher partial waves ($L\geq 1$) partial waves. This approximation is justified from the fact that at the temperatures of interest $T=5-10 \text{ MeV}$, the second virial coefficient coming from p-wave interactions between neutrons is $\sim 2$ orders of magnitude less than the second virial coefficient coming from s-wave interactions. Second, we do not include  partial wave mixing in nucleon-nucleon scattering and we neglect the contribution of protons in the medium and do not include the effect of charged weak currents. Third, we neglect the excitation of more than one particle from the ground state. This can be justified when the typical energy transfer $q_o \simeq q v_{th} \gg \Gamma_n$ where $\Gamma_n$ is the scattering rate of neutrons in the gas. All of these effects can  be included in a straightforward manner, and will be discussed in a follow-up paper.


\section{Calculation}

\begin{figure}[h!]
\includegraphics[scale=0.50]{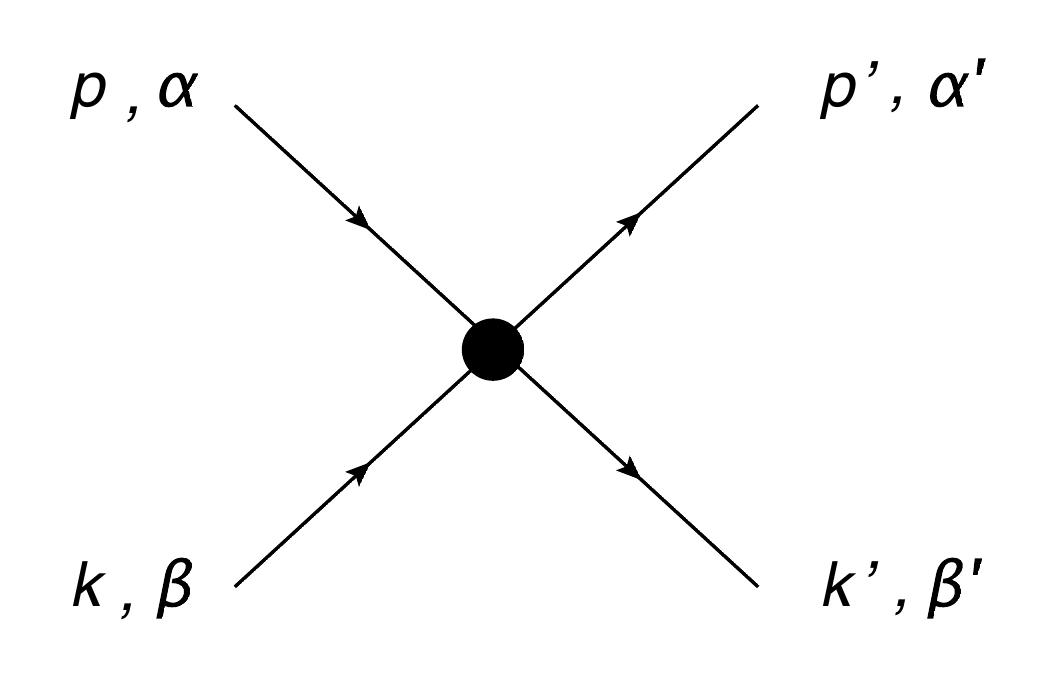}
\caption{The pseudo-potential vertex  equal to $\frac{4\pi}{M} \frac{1}{2}\Big[\frac{\delta(\frac{|p-k|}{2})}{\frac{|p-k|}{2}} + \frac{\delta(\frac{|p'-k'|}{2})}{\frac{|p'-k'|}{2}} \Big](\delta_{\alpha \alpha'}\delta_{\beta \beta'}-\delta_{\alpha \beta'}\delta_{\beta \alpha'})$  }
\label{fig:vertex}
\end{figure}

\noindent As already stated, we compute the dynamic structure factor in the virial expansion. Denoting the contribution to the structure factor $S(q_0,q)$ at $n^{\text{th}}$ order in the fugacity ($z$) expansion as $S_n(q_0,q)$ we can write
\beq
S(q_0,q) = S_1(q_0,q) + S_2(q_0,q)+ S_3(q_0,q) + ...
\label{eq:decomp}
\eeq
 Since the structure factor should reduce to zero in vacuum, the leading nonzero contribution to it appears only in the first order in the virial expansion. As mentioned before, the number of particle-hole loops in a Feynman diagram contributing to the density-density or the spin-spin correlation identifies the lowest order in the virial expansion at which the diagram contributes \cite{Bedaque:2002xy}. This helps fix the diagrams we need to calculate at a given order in the virial expansion. To elaborate further, the first order in virial expansion includes contributions only from a single particle-hole loop, whereas the second order includes contributions from both single as well as double particle-hole loops. Since we are counting only particle-hole loops, any further reference to loops will solely imply particle-hole loops unless mentioned otherwise. We organize our calculation by splitting up the contributions coming from various loops ($m$) at a given order in virial expansion ($n$) denoted as $S_{\text{m-loop},n}$ to write 
\beq
S_1(q_0,q) = S_{\text{1-loop},1}(q_0,q)
\eeq
\beq
S_2(q_0,q) = S_{\text{1-loop},2}(q_0,q) + S_{\text{2-loop},2}(q_0,q) 
\eeq
and so on. 
Each of these terms are computed below for a low density neutron gas. The neutrons are treated as a 2-component spinor field in Matsubara formalism interacting via only two-body forces defined by the pseudo-potential. The free neutron propagator is given by $G_{\alpha \beta}(ip_0,p) = \delta_{\alpha \beta} G(ip_0,p) = \frac{\delta_{\alpha \beta}}{i p_0 - \xi_p}$, where $\alpha,\beta$ indexes the spin. The neutron-neutron vertex is defined in \fig{fig:vertex}. The phase shift $\delta(p)$ appearing in the vertex are the $^1S_0$ channel p-n scattering phase shifts taken from a partial wave analysis carried out by the Theoretical High Energy Physics Group of the Radboud University Nijmegen, and can be found at \href{http://nn-online.org}{http://nn-online.org}. Our computation for the dynamic structure factor can incorporate phase shifts of any form.  
 \noindent To compute the structure factor, we use the fluctuation-dissipation  theorem \cite{0034-4885-29-1-306} that relates the structure factor to the analytic continuation of the Matsubara correlation function
\beq
S(q_0,q) = -\frac{2}{1-\text{e}^{-\beta q_0}}\chi(q_0,q)
\label{eq:Schi}
\eeq
where $\beta = T^{-1}$ and the susceptibility $\chi(q_0,q)$ is defined as
\beq
\chi(q_0,q) = \text{Im} \mathcal{G}(i q_0 \rightarrow q_0 + i 0^+,q).
\label{eq:chi}
\eeq
Here $\mathcal{G}(iq_0,q)$ is the Fourier transform of the Mastubara time ordered correlator $\mathcal{G}(x,\tau) = -\langle T_{\tau}\{\delta n(x,\tau) \delta n(0,0)\}\rangle$. 

\begin{figure}[h!]
\includegraphics[scale=0.50]{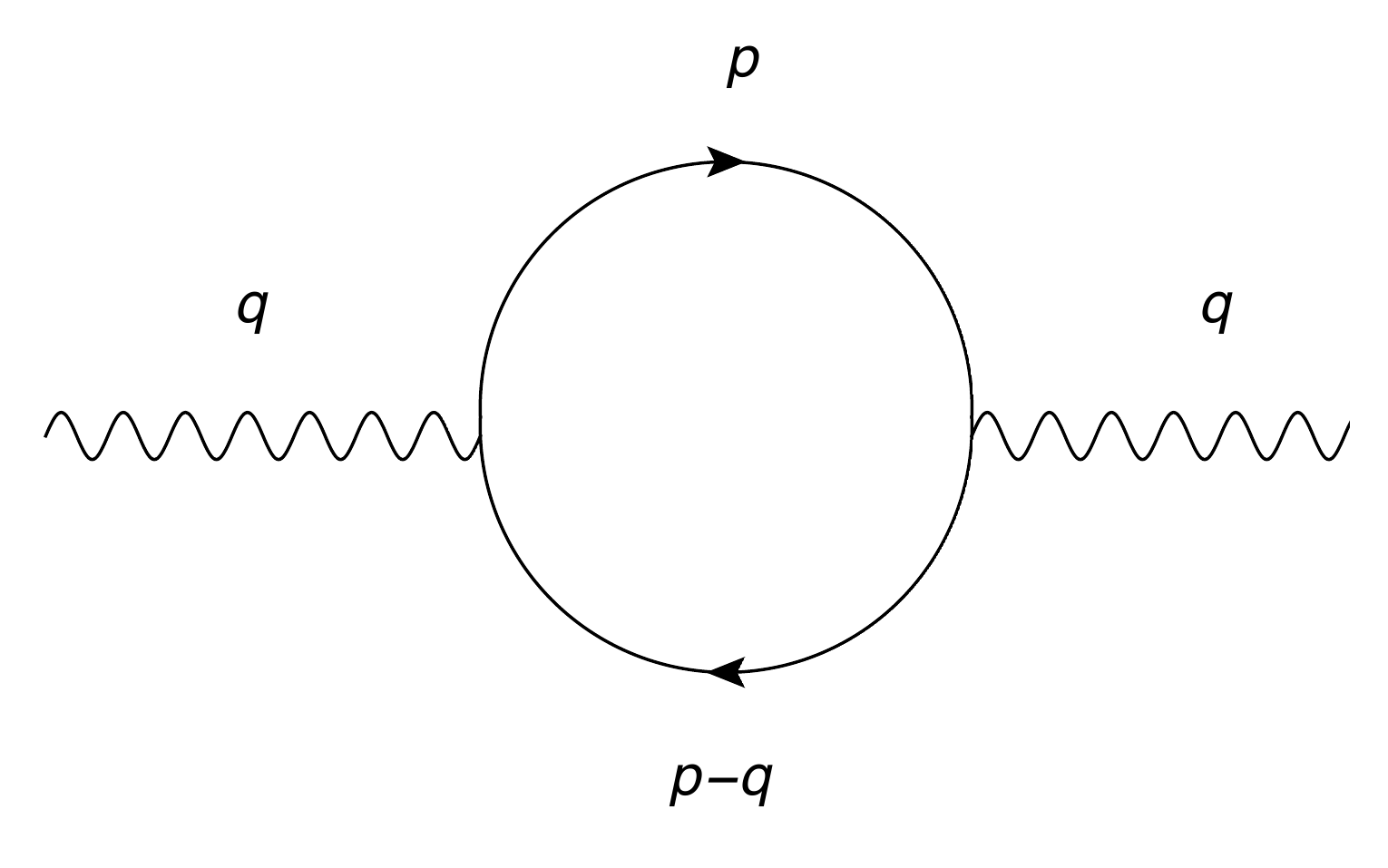}
\caption{Free contribution to the structure factor. This diagram contains both $\mathcal{O}(z)$ and $\mathcal{O}(z^2)$ contributions.}
\label{fig:1-loop}
\end{figure}

\begin{figure}[ht] 
  \begin{subfigure}[b]{0.5\linewidth}
    \centering
    \includegraphics[width=0.8\linewidth]{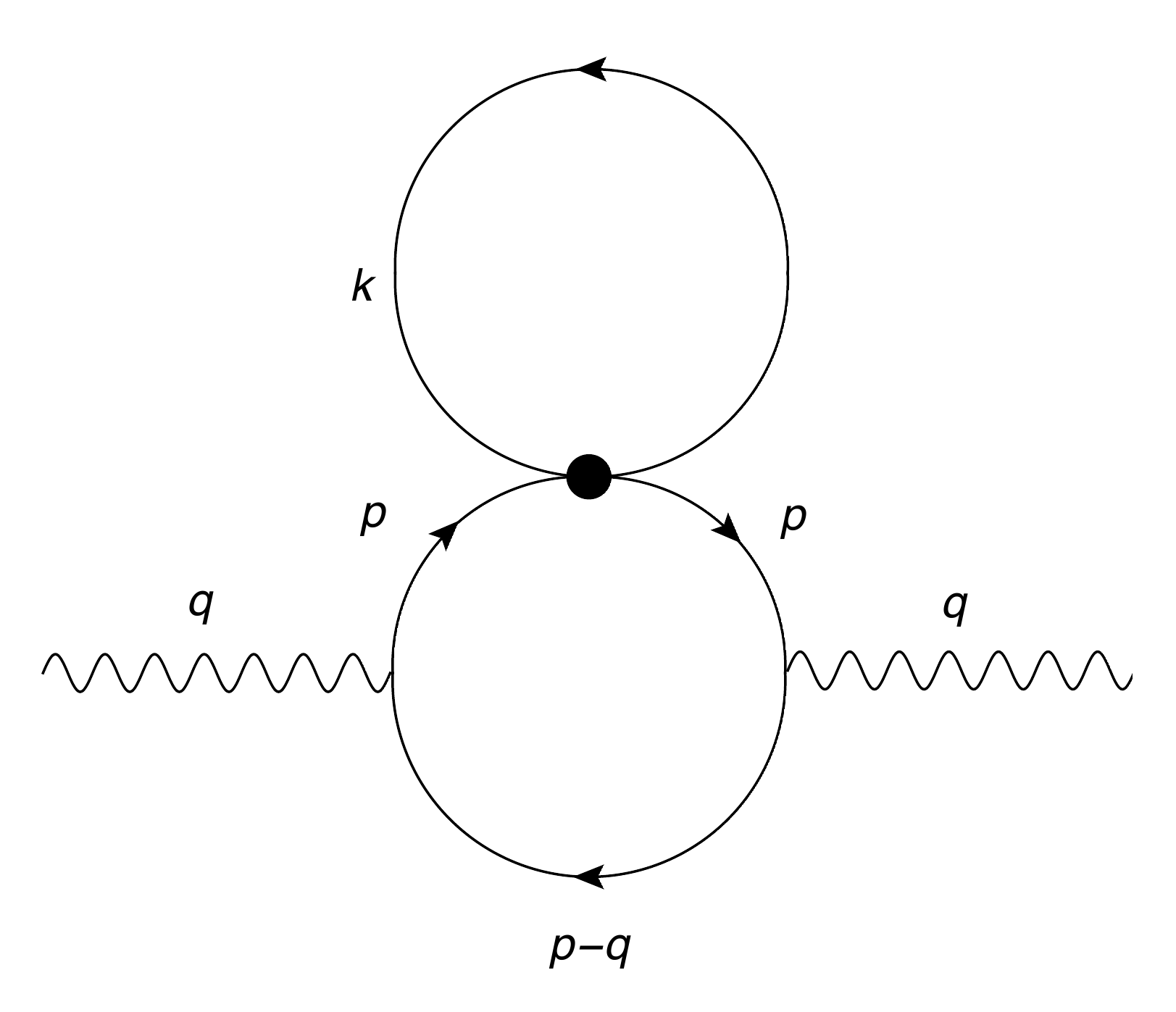} 
    \label{fig7:a} 
    \vspace{4ex}
  \end{subfigure}
  \begin{subfigure}[b]{0.6\linewidth}
    \centering
    \includegraphics[width=0.8\linewidth]{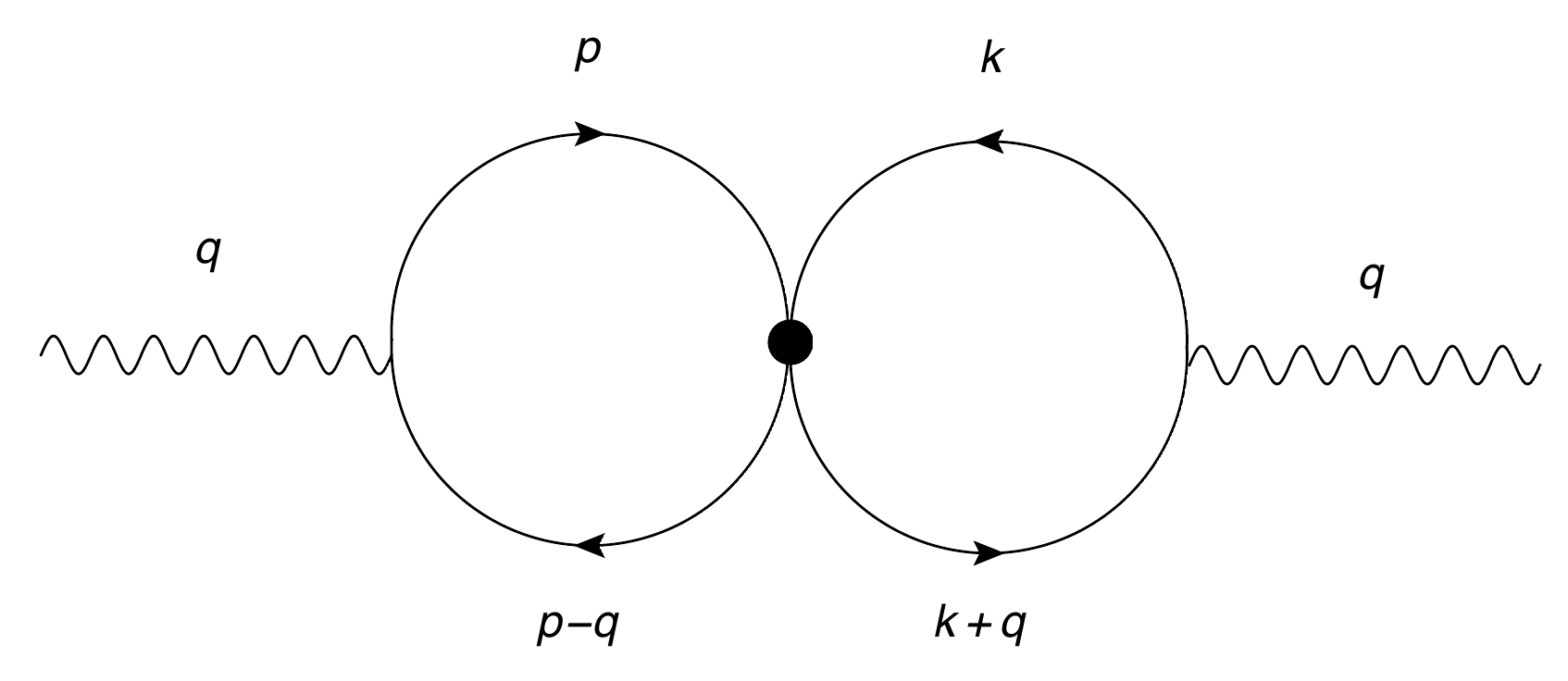} 
    \label{fig7:b} 
    \vspace{4ex} 
    \end{subfigure}
  \caption{Above are the $\mathcal{O}(z^2)$ contributions to the Matsubara correlation function. On the left is the self-energy correction and on the right is the vertex-correction.}
\label{fig:order-z2}
\end{figure}

Diagrams contributing to $\mathcal{G}(iq_0,q)$ up to $\mathcal{O}(z^2)$ are given in Fig. \ref{fig:1-loop} and Fig. \ref{fig:order-z2}.  We separate $\mathcal{G}(iq_0,q)$ into the a-loop and the two 2-loop contributions
\beq\label{eq:G_decompose}
\mathcal{G}(iq_0,q) = \mathcal{G}_{1\text{-loop}}(iq_0,q) + \mathcal{G}_{2\text{-loop},\Sigma}(iq_0,q) 
+ \mathcal{G}_{2\text{-loop}, v}(iq_0,q) + ...
\eeq and similarly for $\chi(q_0,q)$, $S(q_0,q)$ . 

We start by calculating the vector structure function $S_V(q_0,q)$ (we now drop the subscript ``$V$" from $S_V(q_0,q)$ and related functions).
The 1-loop diagram gives
\beq
\mathcal{G}_{1\text{-loop}}(iq_0,q) = 2 T \sum_{p_0}{\int{\dbar^3 p \ G(p)G(p-q)}}
=
2 \int{\dbar^3 k\frac{n(\xi_{k-q/2}) - n(\xi_{k+q/2})}{iq_0 - \frac{k\cdot q}{M}}},
\eeq 
where $\dbar^3 p = d^3p/(2\pi)^3$, $\xi_p = p^2/2M - \mu$, the sum runs over integer multiples of $2\pi T$ and $G(p)$ is the free neutron propagator. 
After using \eq{eq:Schi} and \eq{eq:chi} we obtain the corresponding contribution to the structure funciton

\begin{align}\label{eq:S1}
S_{1\text{-loop}}(q_0,q) & =  -\frac{2}{1-\text{e}^{-\beta q_0}} \Bigg[- \frac{4z}{\lambda^3 q}\sqrt{\frac{\pi M}{2 T}}\text{e}^{ \frac{-\beta M q_0^2}{2 q^2}-\frac{\beta q^2}{8 M}} \text{sinh}(\frac{\beta q_0}{2})-  \Big( z \rightarrow z^2 \text{ and } T \rightarrow \frac{T}{2} \Big)\Bigg] + \mathcal{O}(z^3),
\end{align}
where we have defined the de Broglie wavelength $\lambda = \sqrt{\frac{2\pi}{M T}}$.
As mentioned earlier, $\mathcal{G}_{1\text{-loop}}(iq_0,q)$ contains both $\mathcal{O}(z)$ and
$\mathcal{O}(z^2)$ contributions.

The expression for the 2-loop self-energy diagram (left side of Fig. \ref{fig:order-z2}) is
\beq
\mathcal{G}_{2\text{-loop}, \Sigma}(iq_0,q) = -4 T^2 \sum_{p_0,k_0}{\int{\dbar^3p \dbar^3 k G(p)^2G(k)G(p-q)V(|p-k|, |p-k|)}}
\eeq After using \eq{eq:Schi} and \eq{eq:chi} we find the contribution of the self-energy diagram to the structure function

\begin{align}
&S_{2\text{-loop},\Sigma}(q_0,q) = \frac{-z  }{1-e^{-\beta q_0}}\frac{M e^{\frac{-\beta q^2}{8 M T} }}{\pi q}\Big[2 q_0 \frac{M^2}{q^2}\sinh(\frac{\beta q_0}{2})e^{\frac{-M q_0^2}{2 T q^2}}(\Sigma(\sqrt{M^2 q_0^2/q^2+q^2/4+M q_0})-(q_0\rightarrow -q_0) )\nonumber\\ 
&+ \int_{|\frac{M q_0}{q} |}^{\infty}{dk~k e^{\frac{-\beta k^2}{2 M}}\big\{\beta( \Sigma(\sqrt{k^2+q^2/4+M q_0})e^{-\beta q_0/2}  - (q_0\rightarrow -q_0))}+ \beta \cosh(\beta q_0/2)( \Sigma(\sqrt{k^2+q^2/4-M q_0})- (q_0 \rightarrow - q_0))\nonumber\\
& -M \sinh(\beta q_0 /2)(\frac{\Sigma'(\sqrt{k^2+q^2/4-M q_0})}{\sqrt{k^2+q^2/4-M q_0}}+(q_0 \rightarrow - q_0)\big\}\Big]\label{eq:SSigma}
\end{align} where $\Sigma(p)$ is the self-energy of the neutrons given by
\bea
\Sigma(p)  =  \frac{4 \pi}{M}T \sum_{k_0}{\int{\dbar^3 k G(k)\frac{\delta(|p-k|/2)}{|p-k|/2}}}
=\frac{2 z T}{\pi p}\int_{0}^{\infty}{dk  \Big[\text{e}^{\frac{-\beta(k-p)^2}{2M}}-\text{e}^{\frac{-\beta(k+p)^2}{2M}}\Big] \delta(k/2)} + \mathcal{O}(z^2).
\eea and $\Sigma'(p) = \frac{d}{d p}\Sigma(p)$. Notice that we only need $\Sigma(p)$ compute to  $\mathcal{O}(z)$ as it enters in $S_{2\text{-loop},\Sigma}(q_0,q)$ inside a particle-hole loop.

Finally, the diagram on the right panel of Fig. \ref{fig:order-z2} gives

\bea
\mathcal{G}_{2\text{-loop}, v}(iq_0,q) &=& -2 T^2 \sum_{p_0,k_0}{\int{\dbar^3p \dbar^3 k G(p)G(k)G(p-q)G(k+q) V(|p-k|, |p-k-2q|)}}\\
&=&
\frac{8 \pi}{M} \int{\dbar^3 k\frac{(n(\xi_{k+q/2}) - n(\xi_{k-q/2}))(n(\xi_{p+q/2}) - n(\xi_{p-q/2})}{(iq_0 - \frac{p\cdot q}{M})(iq_0 - \frac{k\cdot q}{M})}\frac{\delta(|k-p+q|/2)}{|k-p+q|/2}}\nn.
\eea Again, using \eq{eq:Schi} and \eq{eq:chi} we find

\bea\label{eq:SV}
 S_{2\text{-loop},v
}(q_0,q)  &=& \frac{-2M T z^2}{1-e^{-\beta q_0}}\frac{e^{\frac{-\beta q^2}{4 M}} }{q \pi^2}
\int_0^\infty \!\!\!dkk^2 e^{-{\frac{\beta k^2}{M}} }
  \int_{-1}^{1}{dx~\frac{M}{2 k q x}\Big[\frac{\delta(\sqrt{k^2 + q^2 /4 + k q x})}{\sqrt{k^2 + q^2 /4 + k q x}}+\frac{\delta(\sqrt{k^2 + q^2 /4 - k q x})}{\sqrt{k^2 + q^2 /4 - k q x}}  \Big]  } \\
&& \times \Bigg[\left(
2 \cosh\left(\beta\left(q_0 - \frac{k q x}{M}\right)\right) e^{\frac{-\beta M}{q^2}(q_0 - \frac{k q x}{M})^2}
-  x\rightarrow -x  \right)
 - \left( 2 \cosh\left(\frac{\beta k q x}{M}\right) e^{\frac{-\beta M}{q^2}(q_0 - \frac{k q x}{M})^2}
 -x\rightarrow -x\right) \Bigg]   \nn.
\eea

We conclude this section by extending these calculations to $S_A(q_0,q)$ (again, we temporarily drop the subscript ``$A$" from $S_A(q_0,q)$  and related functions). Recall that the dynamic structure factor corresponding to spin fluctuations is given by
\beq
S_{i j}(q_0,q) = \int{d^4x \text{e}^{iq_0 t - i q x} \langle \delta s_i(x,t) \delta s_j(0,0) \rangle}
\eeq
where the operator $\delta s_i(x,t)\equiv \psi^{\dagger}(x,t) \sigma_i \psi(x,t)- \langle \psi^{\dagger} \sigma_i \psi \rangle$ and the $\sigma_i$ are the Pauli Matrices. The $^1S_0$ interaction is spin symmetric, so clearly $\langle \delta s_i \delta s_j \rangle \sim \delta_{ij}$. 

The diagrams contributing to the spin-spin correlator are of the same form as the ones in \fig{fig:order-z2} except now there is an insertion of a spin operator on the vertices with a wavy line. The only consequence of this insertions is that the last (vertex correction) diagram acquires an extra minus sign compared to the density-density correlator, thus:

\beq
S_{A,2}(q_0,q) = S_{\text{1-loop},2}(q_0,q) + S_{2\text{-loop},\Sigma}(q_0,q) - S_{2\text{-loop}, v}(q_0,q).
\label{eq:spin-order-z2}
\eeq
The physical interpretation of \eq{eq:spin-order-z2} is apparent: the vertex correction contribution to the spin structure factor is suppressed due to spin anti-alignment in the $^1S_0$ channel. For attractive s-wave interactions, nucleon-nucleon correlations with anti-alignment spin are favored over those in which the spins are aligned. This implies that we can expect the density response to be enhanced,  and correspondingly the spin response to be suppressed.  

The results in Eqs. (\ref{eq:spin-order-z2}),  (\ref{eq:SV}), (\ref{eq:SSigma}) and (\ref{eq:S1}) are our central results. They allow the calculation of the structure factors in terms of two dimensional integrals that are then computed numerically. Below we will provide very good analytic fits to these results that neatly summarize these results.

\section{Sum Rules}
\label{sec:sum-rules} 

As a check on our calculation and to validate the use of the pseudo-potential which is intended to capture the non-perturbative of the ladder summation, will show that sum rules, derived on general grounds, are indeed satisfied by our results. First, the following  thermodynamic sum rule relates the vector structure function to a
thermodynamic quantity \cite{nozieres1999theory}:
\beq
\int_{-\infty}^{\infty}{\frac{dq_0}{2 \pi}~S_V(q_0,q\rightarrow0)}  = T \frac{\partial n}{\partial \mu} =  \frac{2 z}{\lambda^3}(1 + 4b_2z + ...),
\label{eq:density-sum-rule}
\eeq
where $b_2$, the second virial coefficient, is given by the Beth-Uhlenbeck relation \cite{Beth1937915} 
\beq\label{eq:beth-uhlenbeck}
b_2 = -\frac{1}{2^{5/2}}+ \frac{\sqrt{2}}{\pi} \int_{0}^{\infty}{dk~ \frac{d \delta(k)}{dk}\text{e}^{-\frac{\beta k^2}{M}}}.
\eeq where $\delta(k)$ is the phase shift of the ${}^1S_0$ partial wave and $k = |k_1 - k_2|/2$ is the difference in incoming momenta.

The spin structure factor satisfies a similar sum rule \cite{Horowitz:2006pj}:
\beq
\int_{-\infty}^{\infty}{\frac{dq_0}{2 \pi}~S_A(q_0,q\rightarrow0)}  =   \frac{2 z}{\lambda^3}(1 + 4b_{2,free}z + ...),
\label{eq:spin-sum-rule}
\eeq with $b_{2,free}=-2^{-5/2}$.
We verified numerically  that both sides of Eqs. (\ref{eq:density-sum-rule}) and (\ref{eq:spin-sum-rule}) agreed for a array of parameter values and different phase shifts. In addition, our calculations were repeated in the Schwinger-Keldysh formalism which leads to different, but equivalent expressions. These expressions make it easy to see that \eq{eq:density-sum-rule} is satisfied exactly for any phase shift and parameter values (derived in appendix).

A second sum rule that can be used is the so-called \emph{f-sum} rule \cite{nozieres1999theory} which was defined earlier in \eq{eq:FSumGeneral}. Since nuclear interactions conserve baryon number, the interaction commutes with the density operator and 
\beq
\int_{-\infty}^{\infty}{\frac{dq_0}{2\pi}~q_0 S_V(q_0,q) = \frac{q^2}{2M}n}\,. 
\label{eq:f-sum-rule}
\eeq 
In addition, since we only consider s-wave interactions, it also commutes with the spin operator, and of this case the dynamic structure for spin $S_A(q_0,q)$ also satisfies the above sum rule.  We numerically verified that \eq{eq:f-sum-rule} was satisfied for several combinations of parameter values and phase shifts. However, as noted earlier in the discussion pertaining to \eq{eq:FSumGeneral}, the f-sum rule for the spin dynamical structure function does not vanish in the long-wavelength limit when the Hamiltonian contains operators that do not commute with the nucleon spin operator. Such operators enhance the contribution of multi-particle excitations \cite{Raffelt:1993ix,Olsson:2002yu}, and their contribution to the dynamical structure function is necessary to satisfy the f-sum rule in \eq{eq:FSumGeneral}. In this work, since we only include s-wave interactions, it is consistent to neglect these contributions in the long wavelength limit. 


We conclude this section by estimating the range of validity for our virial expansion of the structure factor. 
 The condition for the third term of the virial expansion of $T\partial n/\partial \mu$ to be smaller than the second term is that 
\beq
z < \bigg|\frac{4 b_2}{9 b_3}\bigg|.
\eeq The second virial coefficient for neutrons interacting in the $^1S_0$ channel is nearly constant in the temperature range $T=5-10$ MeV and has a value $b_2 = 0.305$. We can estimate $b_3$ for the neutron gas as being equal to $b_3$ for a dilute fermi gas in the BEC-BCS crossover region, which was computed theoretically in \cite{PhysRevLett.102.160401} to be temperature independent and have a value of $b_3 =-0.291(1)$. This estimate yields the condition for validity of the virial expansion to be
\beq\label{eq:z<0.4}
z \alt 0.47.
\eeq Since the static structure factor at zero momentum transfer is determined by the susceptibility $T\partial n/ \partial\mu$, it is reasonable to expect that the range of validity of the virial expansion of the structure factor is also given by \eq{eq:z<0.4}


\section{Scaling functions and analytical fits}

 \begin{figure}[h!]
\includegraphics[scale=0.7]{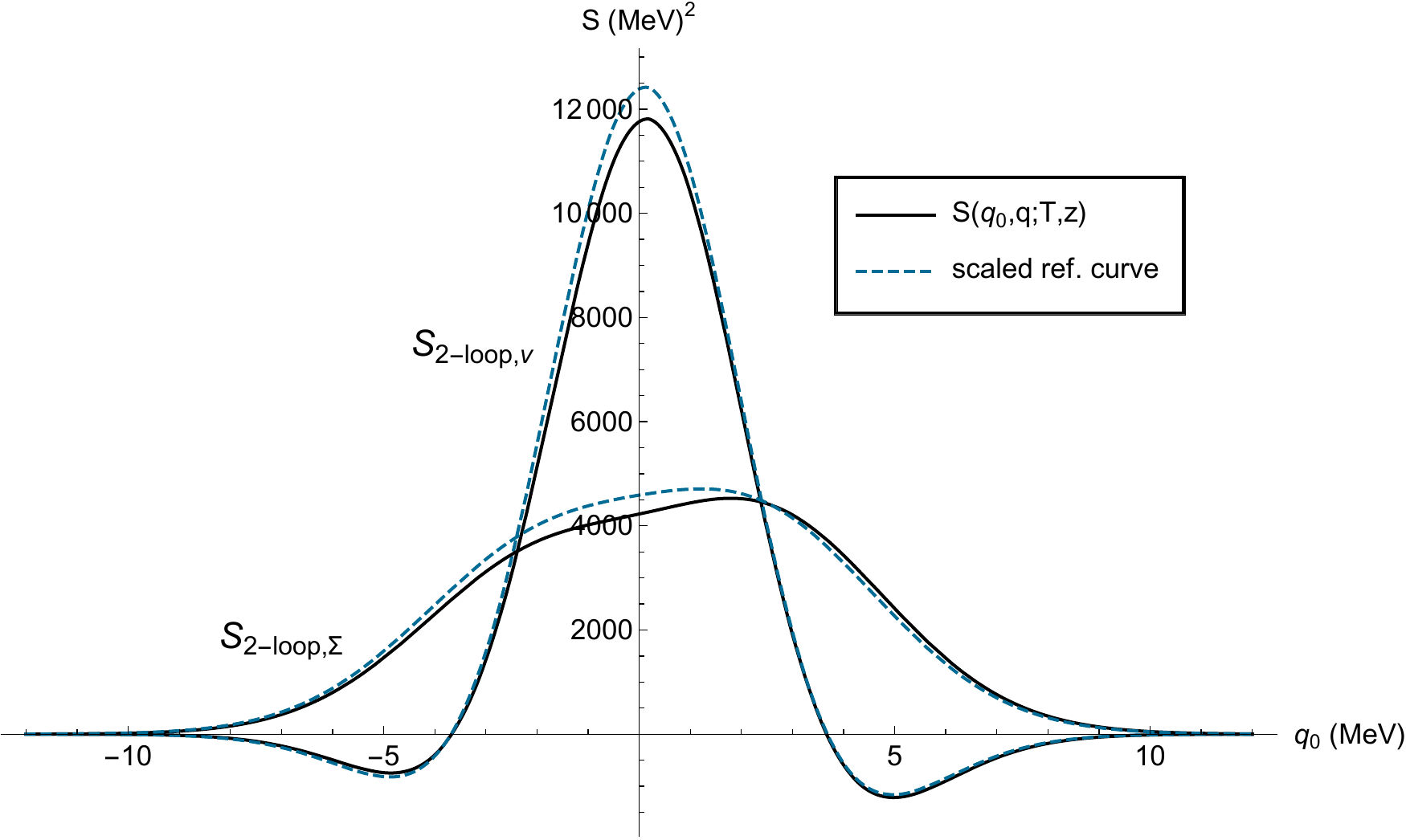}
\caption{ In black are the exact dynamic structure factors of \eq{eq:SSigma} and \eq{eq:SV} at $T = 10$ MeV, $z=1/4$ and $q = 30$ MeV as a function of energy transfer. These $T$ and $q$ represent the upper limits of validity of our scaling law. In dashed blue is the result of applying the scaling law  \eq{eq:S-scaling}  to the reference curves of \eq{eq:S-scaling-function}. The agreement between the exact structure factor and the scaling law strictly improves for lower values of $T$ and $q$.
 }
\label{fig:fit}
\end{figure}

The structure functions $S_A(q_0)$ and $S_V(q_0,q)$ can be written in terms of the functions $S_{1-loop}(q_0,q)$, $S_{2-loop,\Sigma}(q_0,q)$ and $S_{2-loop, V}(q_0,q)$ through the relations

\bea
S_V(q_0,q) &=&  S_{\text{1-loop}}(q_0,q) + S_{2\text{-loop},\Sigma}(q_0,q) + S_{2\text{-loop}, v}(q_0,q), \\
S_A(q_0,q) &=&  S_{\text{1-loop}}(q_0,q) + S_{2\text{-loop},\Sigma}(q_0,q) - S_{2\text{-loop}, v}(q_0,q).
\eea

While $ S_{\text{1-loop},2}(q_0,q) $ has a very explicit form given by \eq{eq:S1}, the expressions for 
$S_{2\text{-loop},\Sigma}(q_0,q)$ and $S_{2\text{-loop}, v}(q_0,q)$ involve two dimensional integrals that need to be computed numerically. It would be useful then to have a more explicit, even if approximate, expression for these functions. To that end, we first notice that they are a function of the temperature $T$, the fugacity $z$ and the energy and momentum transfers $q_0$ and $q$. The dependence on $z$ is, by definition, a factor of $z^2$. We empirically find that there is an approximate scaling relation allowing us to express $S_{2\text{-loop},\Sigma}(q_0,q)$ and $ S_{2\text{-loop}, v}(q_0,q)$ in terms of functions of a single variable: 

\bea\label{eq:S-scaling}
S_{2\text{-loop},\Sigma}(q_0,q;T,z) 
&\approx& 
\frac{z^2}{\bar{z}^2}\frac{1- exp\left(- \beta q_0 \left(\frac{\bar q}{q}\sqrt{\frac{T}{\bar{T}}}\right)\right) }
{1- exp\left(- \beta q_0 \sqrt{\frac{\bar{T}}{T}}\right) }
S_{2\text{-loop},\Sigma}\left(q_0 \frac{\bar q}{q}\sqrt{\frac{\bar{T}}{T}},\bar q;\bar{T},\bar z\right), \\
S_{2\text{-loop}, v}(q_0,q;T,z) 
&\approx&
\frac{z^2}{\bar{z}^2} \frac{1- exp\left(- \beta q_0 \left(\frac{\bar q}{q}\sqrt{\frac{T}{\bar{T}}}\right)\right) }
 {1- exp\left(- \beta q_0 \sqrt{\frac{\bar{T}}{T}}\right) }
S_{2\text{-loop}, v}\left(q_0 \frac{\bar q}{q}\sqrt{\frac{\bar{T}}{T}}, \bar q; \bar{T},\bar z\right),
\eea 
where $\bar q, \bar{T}$ and $\bar z$ are any momentum, temperature and fugacity reference scales. Choosing $\bar q = 1$ MeV, $\bar{T}=5$ MeV and $z=1/4$, the functions 
$S_{2\text{-loop},\Sigma}(q_0, \bar q; \bar{T},\bar z)$ and $S_{2\text{-loop}, v}(q_0, \bar q; \bar{T},\bar z)$ are well parametrized, in the relevant $q=0-30 \text{ MeV}$, $T=5-10 \text{ MeV}$  range of parameters by

\bea\label{eq:S-scaling-function}
S_{2\text{-loop},\Sigma}\left(q_0,\bar q; \bar{T},\bar z\right) &=& A_1~e^{-|\frac{q_0}{\sigma_1}|^{2.75}}, ~A_1 = (262.7\text{ MeV})^2,~ \sigma_1 = 1.252\times10^{-1}\text{ MeV}\\
S_{2\text{-loop}, v}\left(q_0, \bar q; \bar{T},\bar z\right)      &=&  A_2~e^{-\frac{q_0^2}{\sigma_2^2}} \text{cos}\left(\frac{q_0}{\omega}\right), ~~A_2=(430.7\text{ MeV})^2 ,~\sigma_2 = 8.626\times10^{-2}\text{ MeV},~~\omega = 5.560\text{ MeV}\nn
\label{eq:fit}
\eea Note that there is no restriction on $z$ for use of the scaling law since the $z$ dependence is known to be a factor of $z^2$. Fig. \ref{fig:fit} shows how well the full result compares to the scaling functions under the most extreme circumstances. Agreement only improves when the temperature or momentum transfer is decreased. Considering the all the uncertainties involved in our calculation, the use of the analytic expressions
in \eq{eq:fit} are justified in most applications.

\section{Discussion of Structure Factors}

In the following sections we discuss important physical features of the structure functions we computed and the corresponding
neutrino cross sections.

\begin{figure}[h!]
\includegraphics[scale=0.8]{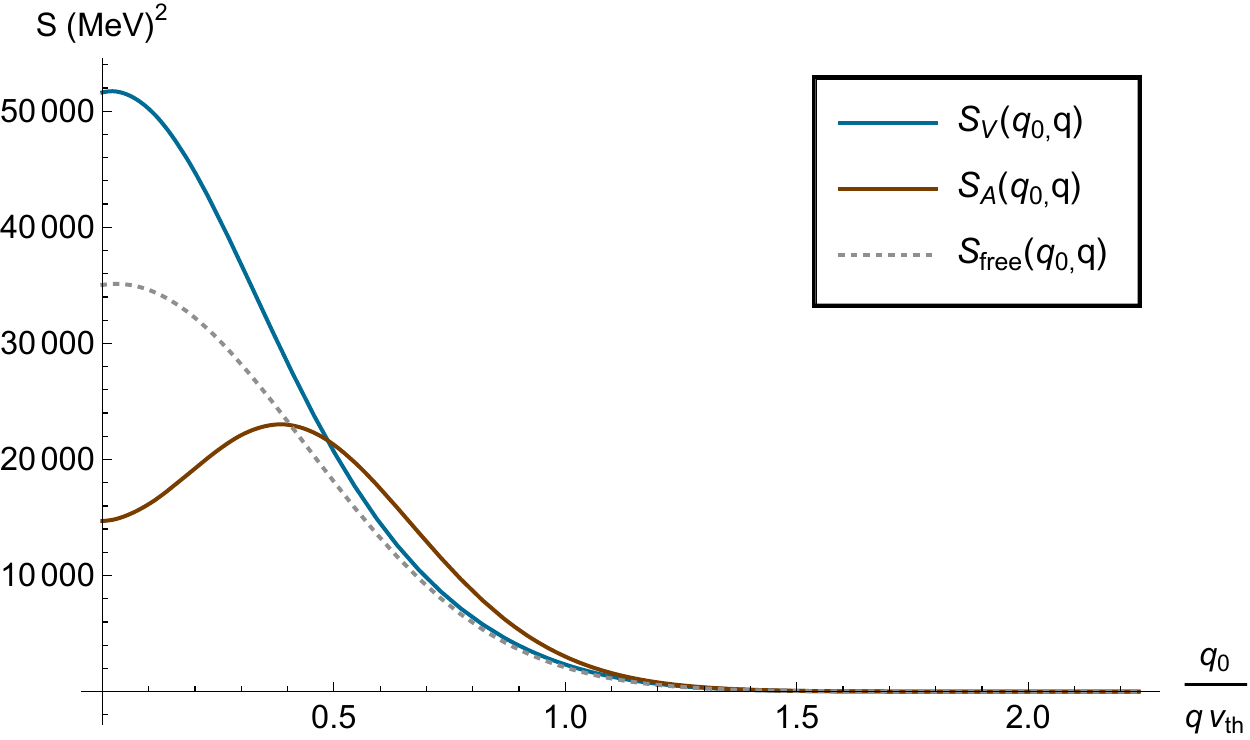}
\caption{Here we demonstrate the dramatic effect that neutron correlations have on the dynamic structure factor. We plot three observables: the dynamic structure factor for density correlations $S_V$ with all contributions up to $\mathcal{O}(z^2)$ in blue, the dynamic structure factor for spin correlations $S_A$ with all contributions up to $\mathcal{O}(z^2)$ in brown, and for comparison we have in dotted grey the free gas density structure factor to $\mathcal{O}(z)$. Here the momentum transfer is chosen to be $q=10 \text{ MeV}$ and we chosen the bulk parameters $T=5 \text{ MeV}$, $z=1/4$ (corresponding to a density of $n/n_{nuc} = 9\times 10^{-3}$).   }
\label{fig:money1}
\end{figure}

The most salient feature to notice in our results, generic in the relevant parameter range, is a substantial enhancement of the density-density correlation and a suppression of the spin-spin correlation.
 In fact, in \fig{fig:money1} we show both the vector and axial structure functions for a free theory and the full result including $\mathcal{O}(z^2)$ correlations for parameters typically present in the neutrinosphere ($T=5 \text{ MeV}$, $z=1/4$, corresponding to a density of $n/n_{nuc} = 9\times 10^{-3}$). This sizeable impact of two body correlations, even at reasonable $z$, can be attributed to the large neutron phase shifts. Of course, at smaller values of $z$, the enhancement/suppression is less pronounced. 

\begin{figure}[h!]
\includegraphics[scale=0.8]{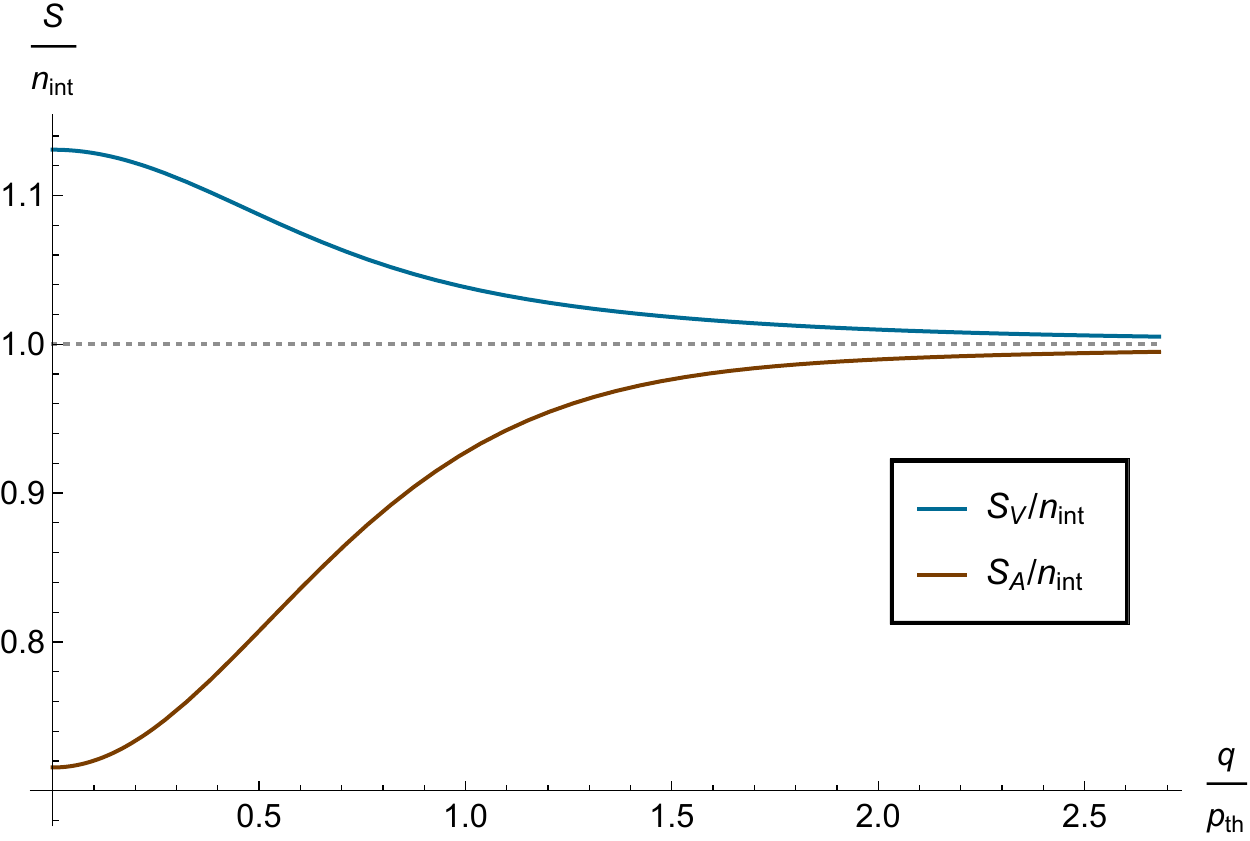}
\caption{Here we show the the static structure factor, computed to $\mathcal{O}(z^2)$, for density (blue) and spin (brown) at the representative temperature $T = 5\text{ MeV}$ and fugacity $z=1/4$ (corresponding to a density of $n/n_{nuc} = 9\times 10^{-3}$). Both curves are normalized by the density computed to $\mathcal{O}(z^2)$ and we plot against the momentum scaled by the thermal momentum $p_{th}\equiv \sqrt{6 M T}$. Once again it is clear that at low momenta, the density response is enhanced while the spin response is suppressed. The convergence of both static structure factors to the density is non-trivial and is predicted by the operator product expansion.}
\label{fig:static-structure-factors}
\end{figure}

The static structure factors are defined by
 \beq
S_V(q) \equiv \int_{-\infty}^{\infty}{\frac{dq_0}{2\pi}~S_V(q_0,q)}, \qquad S_{A}(q) \equiv \int_{-\infty}^{\infty}{\frac{dq_0}{2\pi}~S_A(q_0,q)}
\eeq and are shown in Fig. \ref{fig:static-structure-factors}. The static structure factors are a useful probe of the medium, and have been up until now the only resource for computing neutrino scattering rates through \eq{eq:diff_elas}. We comment on the efficacy of the static structure factor's use in computing neutrino scattering rates (as opposed to the dynamic structure factor) later on. The asymptotic behavior at  large values of the momentum transfer $q$ of both the density-density and spin-spin static structure factors approach the value of the density $n$, as OPE arguments demand \cite{PhysRevX.7.011022}. This convergence is demonstrated in \fig{fig:static-structure-factors} and is analytically demonstrated in the appendix. At small values of the momentum transfer $q$, the static structure factors exhibit the same kind of enhancement (for $S_V$) or suppression (for $S_A$) as the dynamic structure factors, in line with previous observations \cite{Horowitz:2016gul}. 

\section{Results for Neutrino Scattering}

\begin{figure}[h!]
\includegraphics[scale=0.9]{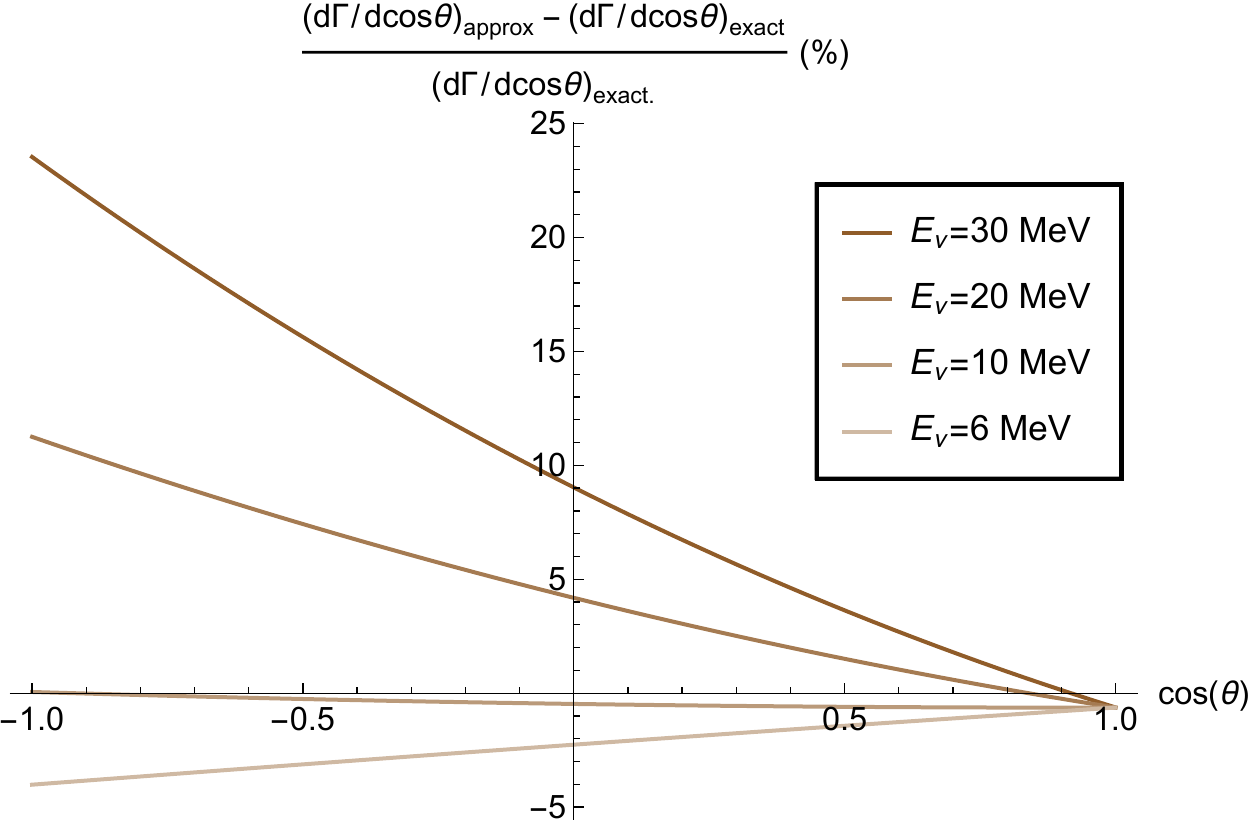}
\caption{To quantify the error incurred on the differential scattering rate using the static approximation \eq{eq:diff_elas}, we plot the difference between the differential scattering rate calculated using the dynamic structure factor \eq{eq:diff2_rate} and the static structure factor \eq{eq:diff_elas}. The former is denoted ``exact" while the latter is denoted ``approx". It is seen that for neutrino energies $E_{\nu}<10$ MeV, scattering rates are systematically under-predicted by no more than $\leq 5 \%$. However, for $E_{\nu}>10$ MeV backscattering quickly becomes wildly overestimated. The thermodynamic parameters are $T=5 \text{ MeV}$ and $z=1/4$. }
\label{fig:compare}
\end{figure}

The neutrino differential scattering rate is determined by the dynamic structure factors through \eq{eq:diff2_rate}.
Since the dynamic structure factor is difficult to compute, it is customary to approximate the scattering rate by utilizing the static structure factor, which is much easier to compute, via \eq{eq:diff_elas}. Now having  a computation of the dynamic structure factors, we can ascertain the impact of this approximation. The comparison between the ``exact" (obtained from the dynamic structure factors) and ``approximate" (obtained from the static structure factors) are shown in \fig{fig:compare}.
We find that the departure of our results from the approximate result is relatively small for smaller neutrino energies ($6$ to $10$ MeV) but is significant for higher neutrino energies around $30$ MeV.

\begin{figure}[h!]
\includegraphics[scale=0.48]{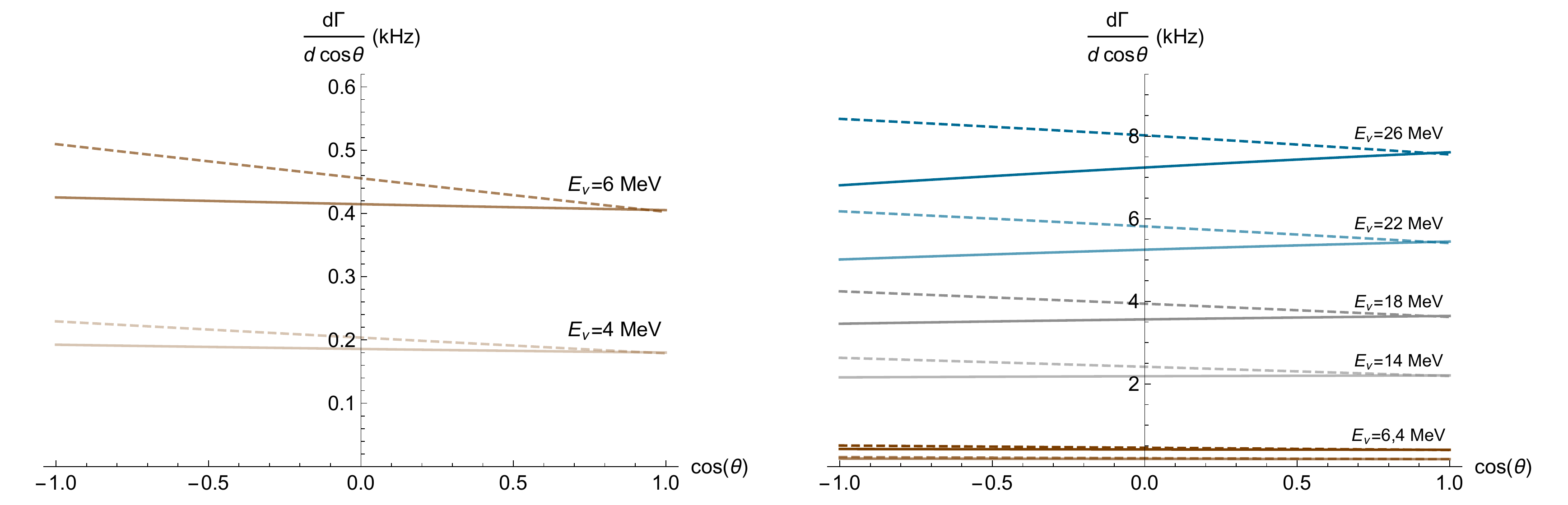}
\caption{Figures show the differential rate for neutrino scattering at $T=5 \text{ MeV}$ and $z=1/4$ over a  range of incoming neutrino energies. The right panel is a zoomed out view of the plot on the left, which focuses on the low energy range.  Dotted lines correspond to the $\mathcal{O}(z)$ free theory predictions, while the solid lines come from the $\mathcal{O}(z^2)$ theory.}
\label{fig:differential-cross-section}
\end{figure}


The consequences of the neutron correlations on neutrino scattering of the are illustrated in Fig. \ref{fig:differential-cross-section}. It is found that the main effect on scattering due to neutron correlations is to strongly suppress back scattering. This can be understood by noting that neutrinos in the ultra-relativistic limit preserve their helicity and thus back scattering can only occur through their axial current coupling to the nucleon spin. However, as is demonstrated in Fig. \ref{fig:static-structure-factors}, $\mathcal{O}(z^2)$ interactions suppress the spin fluctuations, and correlate nearby neutron pairs into spin singlets due to the attractive ${}^1S_0$ interactions. In contrast, the enhancement of the density fluctuations 

In order to make it easier to use the results in \eq{eq:S-scaling-function} and \eq{eq:S-scaling} 
it would be useful to have a simple expression relating the fugacity $z$ to the neutron density $n$. Up to $\mathcal{O}(z^2)$ the relation is
\beq
n = \frac{2z}{\lambda^3} (1+2z b_2(T) + \cdots),
\eeq where $b_2$ is given by the Beth-Uhlenbeck formula \eq{eq:beth-uhlenbeck}. Using the neutron-neutron s-wave phase shifts the second virial coefficient $b_2(T)$ is well parametrized by
\beq
b_2(T) = a_0 + a_1 T + a_2 T^2 + \cdots
\eeq with

\beq
a_0 = 0.306, \qquad
a_1 = -1.17 \times 10^{-4}~\text{MeV}^{-1}, \qquad
a_2 = -1.93 \times 10^{-4}~ \text{MeV}^{-2}.
\eeq

\section{Conclusion}

In this work, we examined the effects of neutron interactions on neutrino scattering rates in the neutrinosphere. Although it is difficult to analyze neutrino scattering off cold dense matter in a systematic way due to the absence of a small expansion parameter, in the high temperature dilute gas of the neutrinosphere the fugacity is a small parameter and therefore calculations are much more tractable. The expansion in fugacity is known as the virial expansion. We compute the dynamic structure factor for both density and spin correlations in the virial expansion and extract from these structure factors medium modified scattering rates.  Our work is meant to improve on the previous calculations of neutrino scattering in hot and dilute matter, where the scattering rates are computed in the long wavelength limit and medium effects can be expressed in terms of the equation of state. Though model independent, the long wavelength limit has its limitations because the momentum dependence of observables is completely disregarded. We compute, for the first time, the dependence of the structure factor on energy and momentum transfer from the neutrinos to the medium. We model the neutron-neutron interaction with a pseudo-potential vertex in the ${}^1S_0$ channel. The pseudo-potential approach takes in as input on-shell scattering phase shifts and outputs, upon calculation of Feynman diagrams, dynamical correlations. We find that upon inclusion of two-body correlations, neutrino scattering is suppressed in the medium. In particular, back scattering is most strongly suppressed. Since $^1S_0$  interactions between neutrons  tend to anti-correlate spins into spin 0 singlets and suppress the axial response, backscattering, which can only proceed via the axial current coupling for ultra-relativistic neutrinos $(m_{\nu}/E \rightarrow 0)$, is correspondingly suppressed. Both vector and axial currents contribute to scattering at forward angles, and the modest enhancement of the vector response partially compensates for the suppression of axial response

We have demonstrated that the pseudo-potential model behaves sensibly. In particular, we have shown that the dynamic structure factor extracted from the pseudo-potential approach reproduces \emph{exactly} the thermodynamics of the neutron gas and satisfies the f-sum rule. Additionally, the pseudo-potential reproduces the high momentum predictions for the static structure factor from the operator product expansion. There are several improvements that warrant further study, and we aim to include: (i) higher partial waves; (ii) two particle excitations above the ground state in future work. In addition, to account for short-distance dynamics, two-body currents need to be included consistently.  To access higher densities, the pseudo-potential will need to be replaced either realistic interactions where in the particle-particle channels are summed to higher order or by effective interactions that properly account for the effects of Pauli-blocking and nucleon-self energies in the intermediate states. Although these improvements are warranted, the results presented here already marks an advances over earlier work where corrections due to strong interactions were only included in the static, long-wavelength limit.

\section{Acknowledgments}
N.C.W. and P.B. are supported by U.S. Department of Energy under Contract Number DE-FG02-93ER-40762. S.R. and S.S supported by U.S. Department of Energy under grant Contract Number DE-FG02-00ER41132.


\appendix
\section{Analytical Verification of sum rules}
We have used the Matsubara imaginary time formalism in the calculations presented in the main part of this paper. 
In order to check some results we repeated many of them in the Schwinger-Kelydsh real time formalism.
The real time formalism frequently provides equivalent, but different, expressions for some results that sometimes are easier to interpret physically. 


As a check on our results, we computed $S_V(q_0,q)$ and $S_A(q_0,q)$ in both formalisms and we found exact agreement. As such we are justified in proving the sum rules in either formalism, and so we choose to do so with Schwinger-Keldysh. The strength of the Schwinger-Keldysh approach is the transparency with which the sum rules are demonstrated. 

In the real time formalism every field $\psi$ in the theory is represented by two fields $\psi_+$ and $\psi_-$. Propagators are $2\times 2$ matrices:
\beq
D(p) =
\begin{pmatrix}
D_{++}(p)   &   D_{+-} \\
D_{-+}(p)   &  D_{--}
\end{pmatrix}
=
\begin{pmatrix}
\frac{1}{p_0-\xi_p+i0^{+}}+ 2\pi i n(p_0)\delta(p_0 -\xi_p)    &   2\pi i n(p_0) \delta(p_0-\xi_p) \\
2\pi i (n(p_0) -1) \delta(p_0-\xi_p)    &  - \frac{1}{p_0-\xi_p-i0^{+}}+ 2\pi i n(p_0)\delta(p_0 -\xi_p)    
\end{pmatrix}. 
\eeq The vertices are also doubled but remain diagonal in the $\pm$ space:
\beq
V_+(p) = i \frac{4\pi}{M} \left (\frac{\delta(p)}{p} + \frac{\delta(p')}{p'} \right), 
\qquad 
V_-(p) = - i \frac{4\pi}{M} \left (\frac{\delta(p)}{p} + \frac{\delta(p')}{p'} \right).
\eeq The structure factor, in terms of the fields $\psi_+, \psi_-$, is given by
\bea
S_V(q_0,q) = \int{d^4 x \langle T_c\{\psi^{\dagger}_{-}(x,t) \psi_{-}(x,t) \psi^{\dagger}_{+}(0,0) \psi_{+}(0,0)\}\rangle} \nn\\
S_A(q_0,q) = \int{d^4 x \langle T_c\{\psi^{\dagger}_{-}(x,t) \sigma_3\psi_{-}(x,t) \psi^{\dagger}_{+}(0,0)\sigma_3 \psi_{+}(0,0)\}\rangle}
\eea where $T_c$ is time ordering along the Schwinger Kelydsh contour. A straightforward calculation leads to the expressions

\bea\label{eq:SK}
S_{2\text{-loop},\Sigma}(q_0,q) 
&=&
8\pi^2 \int{\dbar^4 p \Sigma(p)\delta'(p_0 -\xi_p)\Big[n(p_0 - q_0)\delta(p_0-q_0-\xi_{p-q}) + n(p_0)\delta(p_0 + q_0 -\xi_{p+q})\Big]} \\
S_{2\text{-loop},v}(q_0,q)
&=&
\frac{16\pi^2 z^2}{M}e^{\frac{-\beta q^2}{4 M}}\int \dbar^3 p \dbar^3 k\ \delta(q_0-k\cdot q/M)
e^{-\beta(\epsilon_p + \epsilon_k)}
P\left(\frac{1}{q_0 - p\cdot q/M}\right) 
\left(e^{-\frac{\beta(p-k)\cdot q}{2M}}-e^{\frac{\beta(p+k)\cdot q}{2M}}\right)   \nn\\
& &\times V\left(\left|\frac{k-p-q}{2}\right| ,  \left|\frac{k-p+q}{2} \right| \right)  ,
\eea
where $\delta'(p_0-\xi_p) = \frac{d}{dp_0} \delta(p_0 - \xi_p)$, $P(\frac{1}{x})$ denotes the principal value, $\epsilon_p = p^2/2M$ and $n(p_0) = (e^{\beta p_0}+1)^{-1}$ is the Fermi-Dirac distribution. We first demonstrate the thermodynamic sum rule, which is obtained by integrating over $q_0$ then taking the $q\rightarrow 0$ limit:
\bea
S_{2\text{-loop},\Sigma}(q) &=&
 \int{\frac{dq_0}{2 \pi} S_{2\text{-loop},\Sigma}(q_0,q)} = 4 \pi \int{\dbar^4 p \Sigma(p)\delta'(p_0 -\xi_p)\Big[n(\xi_{p-q}) + n(p_0)\Big]} \nn\\
 &=& 
  2z \int{\dbar^3 p \frac{\Sigma(p)}{T} e^{-\beta \epsilon_p}} + \mathcal{O}(z^3) 
 =
  \frac{4z^2}{\lambda^3}\frac{\sqrt{2}}{\pi}\int_{0}^{\infty}{dk~e^{-\frac{\beta k^2}{M}}\frac{d \delta}{d k}}
  + \mathcal{O}(z^3)  \nn\\
  &=&
  \frac{4 z^2}{\lambda^3}(b_2 - b_{2,free}) + \mathcal{O}(z^3) .
\eea
Note that $S_{2\text{-loop},\Sigma}(q)$ is actually independent of q. Given that $S_V(q\rightarrow 0) = \partial n/\partial\mu = 2\lambda^{-3}z(1+4b_2z+...)$ we see that the self energy diagram contributes a half of the thermodynamic sum rule. The remaining half comes from the vertex. To show this, integrate \eq{eq:SK} over frequencies:
\bea
S_{2\text{-loop},v}(q)
&=&\frac{8\pi z^2}{M}e^{\frac{-\beta q^2}{4 M}}\int \dbar^3 p \dbar^3 k e^{-\beta(\epsilon_p + \epsilon_k)}
P\left(\frac{M}{(k - p)\cdot q} \right) \big[e^{-\frac{\beta(p-k)\cdot q}{2M}}-e^{\frac{\beta(p+k)\cdot q}{2M}}\big]
\nn\\
&&\qquad  \times V\left( \left|\frac{k-p-q}{2} \right|, \left|\frac{k-p+q}{2} \right| \right)     
\eea
Choosing center of mass coordinates $P = k+p$, $K = (k-p)/2$ and letting $q$ approach zero, one finds
\beq
S_{2\text{-loop},v}(q\rightarrow 0)=-\frac{8\pi z^2}{M}\frac{2\sqrt{2}}{\lambda^3} \int{\dbar^3 K\
 e^{-\beta K^2/M}P\left(\frac{M}{K\cdot q} \right) e^{-\frac{\beta K\cdot q}{M}} \frac{\delta(K)}{K} }.
\eeq
Utilizing the identity $\lim_{\alpha\rightarrow 0}\int_{-\alpha}^{\alpha}{d\xi P(\frac{1}{\xi}) e^{-\xi}} = -2 \alpha + \mathcal{O}(\alpha^2)$ one finds
\beq
S_{2\text{-loop},v}(q\rightarrow 0) = \frac{4z^2}{\lambda^3}\frac{\sqrt{2}}{\pi}\int_{0}^{\infty}{dk~e^{-\frac{\beta k^2}{M}}\frac{d \delta}{d k}} = S_{2\text{-loop},\Sigma}(q\rightarrow 0)
\eeq
Thus the thermodynamic sum rule for $S_V(q)$ is verified. Moreover, from the fact that $S_{2\text{-loop},v}(q\rightarrow 0) = S_{2\text{-loop},\Sigma}(q\rightarrow 0)$, one immediately verifies the thermodynamic sum rule for the spin structure factor, $S_A(q\rightarrow 0) = 2\lambda^{-3}z(1+4b_{2,free}+...)$.  

The asymptotic behavior of the structure functions at high momentum shown in Fig. \ref{fig:static-structure-factors} can be obtained analytically. In fact, 
\beq
S_{2\text{-loop},v}(q\rightarrow \infty)=-\frac{8\pi z^2}{M}\frac{2\sqrt{2}}{\lambda^3} \frac{\delta(q)}{q} e^{\frac{-\beta q^2}{4 M}} \int{\dbar^3 K e^{-\beta K^2/M}P(\frac{M}{K\cdot q}) e^{-\frac{\beta K\cdot q}{M}}} 
\eeq
As $q\rightarrow \infty$ the angular integral converges to 
\beq
\int_{-1}^{1}{dx~P(\frac{M}{K q x})e^{-\frac{\beta K q}{M}x}} \rightarrow T \frac{M^2}{K^2 q^2}(e^{\frac{\beta K q}{M}}-e^{-\frac{\beta K q}{M}})
\eeq
Dropping unnecessary numerical factors, one finds
\beq
S_{2\text{-loop},v}(q\rightarrow \infty)\propto \frac{\delta(q)}{q^3} \int{dK~e^{-\beta(K+q/2)^2/M}-e^{-\beta(K-q/2)^2/M}} \rightarrow \frac{\delta(q)}{q^3} \sqrt{M T} \rightarrow 0.
\eeq On the other hand as
$S_{2\text{-loop},\Sigma}(q) = 4 z^2 \lambda^{-3}(b_2 - b_{2,free})$, we see that  
\beq
S_V(q \rightarrow \infty) =  2\frac{z}{\lambda^3}(1+2b_2z+...) = n,
\eeq as depicted in Fig. \ref{fig:static-structure-factors}.

\bibliography{neutrino_medium}

\begin{thebibliography}{21}%
\makeatletter
\providecommand \@ifxundefined [1]{%
 \@ifx{#1\undefined}
}%
\providecommand \@ifnum [1]{%
 \ifnum #1\expandafter \@firstoftwo
 \else \expandafter \@secondoftwo
 \fi
}%
\providecommand \@ifx [1]{%
 \ifx #1\expandafter \@firstoftwo
 \else \expandafter \@secondoftwo
 \fi
}%
\providecommand \natexlab [1]{#1}%
\providecommand \enquote  [1]{``#1''}%
\providecommand \bibnamefont  [1]{#1}%
\providecommand \bibfnamefont [1]{#1}%
\providecommand \citenamefont [1]{#1}%
\providecommand \href@noop [0]{\@secondoftwo}%
\providecommand \href [0]{\begingroup \@sanitize@url \@href}%
\providecommand \@href[1]{\@@startlink{#1}\@@href}%
\providecommand \@@href[1]{\endgroup#1\@@endlink}%
\providecommand \@sanitize@url [0]{\catcode `\\12\catcode `\$12\catcode
  `\&12\catcode `\#12\catcode `\^12\catcode `\_12\catcode `\%12\relax}%
\providecommand \@@startlink[1]{}%
\providecommand \@@endlink[0]{}%
\providecommand \url  [0]{\begingroup\@sanitize@url \@url }%
\providecommand \@url [1]{\endgroup\@href {#1}{\urlprefix }}%
\providecommand \urlprefix  [0]{URL }%
\providecommand \Eprint [0]{\href }%
\providecommand \doibase [0]{http://dx.doi.org/}%
\providecommand \selectlanguage [0]{\@gobble}%
\providecommand \bibinfo  [0]{\@secondoftwo}%
\providecommand \bibfield  [0]{\@secondoftwo}%
\providecommand \translation [1]{[#1]}%
\providecommand \BibitemOpen [0]{}%
\providecommand \bibitemStop [0]{}%
\providecommand \bibitemNoStop [0]{.\EOS\space}%
\providecommand \EOS [0]{\spacefactor3000\relax}%
\providecommand \BibitemShut  [1]{\csname bibitem#1\endcsname}%
\let\auto@bib@innerbib\@empty
\bibitem [{\citenamefont {Burrows}\ \emph {et~al.}(2006)\citenamefont
  {Burrows}, \citenamefont {Reddy},\ and\ \citenamefont
  {Thompson}}]{Burrows:2004vq}%
  \BibitemOpen
  \bibfield  {author} {\bibinfo {author} {\bibfnamefont {A.}~\bibnamefont
  {Burrows}}, \bibinfo {author} {\bibfnamefont {S.}~\bibnamefont {Reddy}}, \
  and\ \bibinfo {author} {\bibfnamefont {T.~A.}\ \bibnamefont {Thompson}},\
  }\href {\doibase 10.1016/j.nuclphysa.2004.06.012} {\bibfield  {journal}
  {\bibinfo  {journal} {Nucl. Phys.}\ }\textbf {\bibinfo {volume} {A777}},\
  \bibinfo {pages} {356} (\bibinfo {year} {2006})},\ \Eprint
  {http://arxiv.org/abs/astro-ph/0404432} {arXiv:astro-ph/0404432 [astro-ph]}
  \BibitemShut {NoStop}%
\bibitem [{\citenamefont {Sawyer}(1975)}]{Sawyer:1975}%
  \BibitemOpen
  \bibfield  {author} {\bibinfo {author} {\bibfnamefont {R.~F.}\ \bibnamefont
  {Sawyer}},\ }\href {\doibase 10.1103/PhysRevD.11.2740} {\bibfield  {journal}
  {\bibinfo  {journal} {Phys. Rev. D}\ }\textbf {\bibinfo {volume} {11}},\
  \bibinfo {pages} {2740} (\bibinfo {year} {1975})}\BibitemShut {NoStop}%
\bibitem [{\citenamefont {Iwamoto}\ and\ \citenamefont
  {Pethick}(1982)}]{Iwamoto:1982zp}%
  \BibitemOpen
  \bibfield  {author} {\bibinfo {author} {\bibfnamefont {N.}~\bibnamefont
  {Iwamoto}}\ and\ \bibinfo {author} {\bibfnamefont {C.~J.}\ \bibnamefont
  {Pethick}},\ }\href {\doibase 10.1103/PhysRevD.25.313} {\bibfield  {journal}
  {\bibinfo  {journal} {Phys. Rev.}\ }\textbf {\bibinfo {volume} {D25}},\
  \bibinfo {pages} {313} (\bibinfo {year} {1982})}\BibitemShut {NoStop}%
\bibitem [{\citenamefont {Horowitz}\ and\ \citenamefont
  {Wehrberger}(1991)}]{Horowitz:1990it}%
  \BibitemOpen
  \bibfield  {author} {\bibinfo {author} {\bibfnamefont {C.~J.}\ \bibnamefont
  {Horowitz}}\ and\ \bibinfo {author} {\bibfnamefont {K.}~\bibnamefont
  {Wehrberger}},\ }\href {\doibase 10.1016/0375-9474(91)90745-R} {\bibfield
  {journal} {\bibinfo  {journal} {Nucl. Phys.}\ }\textbf {\bibinfo {volume}
  {A531}},\ \bibinfo {pages} {665} (\bibinfo {year} {1991})}\BibitemShut
  {NoStop}%
\bibitem [{\citenamefont {Burrows}\ and\ \citenamefont
  {Sawyer}(1998)}]{Burrows:1998cg}%
  \BibitemOpen
  \bibfield  {author} {\bibinfo {author} {\bibfnamefont {A.}~\bibnamefont
  {Burrows}}\ and\ \bibinfo {author} {\bibfnamefont {R.~F.}\ \bibnamefont
  {Sawyer}},\ }\href {\doibase 10.1103/PhysRevC.58.554} {\bibfield  {journal}
  {\bibinfo  {journal} {Phys. Rev.}\ }\textbf {\bibinfo {volume} {C58}},\
  \bibinfo {pages} {554} (\bibinfo {year} {1998})},\ \Eprint
  {http://arxiv.org/abs/astro-ph/9801082} {arXiv:astro-ph/9801082 [astro-ph]}
  \BibitemShut {NoStop}%
\bibitem [{\citenamefont {Reddy}\ \emph {et~al.}(1999)\citenamefont {Reddy},
  \citenamefont {Prakash}, \citenamefont {Lattimer},\ and\ \citenamefont
  {Pons}}]{Reddy:1998hb}%
  \BibitemOpen
  \bibfield  {author} {\bibinfo {author} {\bibfnamefont {S.}~\bibnamefont
  {Reddy}}, \bibinfo {author} {\bibfnamefont {M.}~\bibnamefont {Prakash}},
  \bibinfo {author} {\bibfnamefont {J.~M.}\ \bibnamefont {Lattimer}}, \ and\
  \bibinfo {author} {\bibfnamefont {J.~A.}\ \bibnamefont {Pons}},\ }\href
  {\doibase 10.1103/PhysRevC.59.2888} {\bibfield  {journal} {\bibinfo
  {journal} {Phys. Rev.}\ }\textbf {\bibinfo {volume} {C59}},\ \bibinfo {pages}
  {2888} (\bibinfo {year} {1999})},\ \Eprint
  {http://arxiv.org/abs/astro-ph/9811294} {arXiv:astro-ph/9811294 [astro-ph]}
  \BibitemShut {NoStop}%
\bibitem [{\citenamefont {Horowitz}\ and\ \citenamefont
  {Schwenk}(2006{\natexlab{a}})}]{Horowitz:2005zv}%
  \BibitemOpen
  \bibfield  {author} {\bibinfo {author} {\bibfnamefont {C.~J.}\ \bibnamefont
  {Horowitz}}\ and\ \bibinfo {author} {\bibfnamefont {A.}~\bibnamefont
  {Schwenk}},\ }\href {\doibase 10.1016/j.physletb.2006.05.055} {\bibfield
  {journal} {\bibinfo  {journal} {Phys. Lett.}\ }\textbf {\bibinfo {volume}
  {B638}},\ \bibinfo {pages} {153} (\bibinfo {year} {2006}{\natexlab{a}})},\
  \Eprint {http://arxiv.org/abs/nucl-th/0507064} {arXiv:nucl-th/0507064
  [nucl-th]} \BibitemShut {NoStop}%
\bibitem [{\citenamefont {Horowitz}\ and\ \citenamefont
  {Schwenk}(2006{\natexlab{b}})}]{Horowitz:2006pj}%
  \BibitemOpen
  \bibfield  {author} {\bibinfo {author} {\bibfnamefont {C.~J.}\ \bibnamefont
  {Horowitz}}\ and\ \bibinfo {author} {\bibfnamefont {A.}~\bibnamefont
  {Schwenk}},\ }\href {\doibase 10.1016/j.physletb.2006.09.042} {\bibfield
  {journal} {\bibinfo  {journal} {Phys. Lett.}\ }\textbf {\bibinfo {volume}
  {B642}},\ \bibinfo {pages} {326} (\bibinfo {year} {2006}{\natexlab{b}})},\
  \Eprint {http://arxiv.org/abs/nucl-th/0605013} {arXiv:nucl-th/0605013
  [nucl-th]} \BibitemShut {NoStop}%
\bibitem [{\citenamefont {Mahan}(1993)}]{Mahan}%
  \BibitemOpen
  \bibfield  {author} {\bibinfo {author} {\bibfnamefont {G.~D.}\ \bibnamefont
  {Mahan}},\ }\href@noop {} {\emph {\bibinfo {title} {{Many-Particle
  Physics}}}},\ \bibinfo {edition} {2nd}\ ed.\ (\bibinfo  {publisher}
  {Plenum},\ \bibinfo {address} {New York, N.Y.},\ \bibinfo {year}
  {1993})\BibitemShut {NoStop}%
\bibitem [{\citenamefont {Olsson}\ and\ \citenamefont
  {Pethick}(2002)}]{Olsson:2002yu}%
  \BibitemOpen
  \bibfield  {author} {\bibinfo {author} {\bibfnamefont {E.}~\bibnamefont
  {Olsson}}\ and\ \bibinfo {author} {\bibfnamefont {C.~J.}\ \bibnamefont
  {Pethick}},\ }\href {\doibase 10.1103/PhysRevC.66.065803} {\bibfield
  {journal} {\bibinfo  {journal} {Phys. Rev.}\ }\textbf {\bibinfo {volume}
  {C66}},\ \bibinfo {pages} {065803} (\bibinfo {year} {2002})},\ \Eprint
  {http://arxiv.org/abs/astro-ph/0208453} {arXiv:astro-ph/0208453 [astro-ph]}
  \BibitemShut {NoStop}%
\bibitem [{\citenamefont {Raffelt}\ and\ \citenamefont
  {Seckel}(1995)}]{Raffelt:1993ix}%
  \BibitemOpen
  \bibfield  {author} {\bibinfo {author} {\bibfnamefont {G.}~\bibnamefont
  {Raffelt}}\ and\ \bibinfo {author} {\bibfnamefont {D.}~\bibnamefont
  {Seckel}},\ }\href {\doibase 10.1103/PhysRevD.52.1780} {\bibfield  {journal}
  {\bibinfo  {journal} {Phys. Rev.}\ }\textbf {\bibinfo {volume} {D52}},\
  \bibinfo {pages} {1780} (\bibinfo {year} {1995})},\ \Eprint
  {http://arxiv.org/abs/astro-ph/9312019} {arXiv:astro-ph/9312019 [astro-ph]}
  \BibitemShut {NoStop}%
\bibitem [{\citenamefont {Lykasov}\ \emph {et~al.}(2008)\citenamefont
  {Lykasov}, \citenamefont {Pethick},\ and\ \citenamefont
  {Schwenk}}]{Lykasov:2008yz}%
  \BibitemOpen
  \bibfield  {author} {\bibinfo {author} {\bibfnamefont {G.~I.}\ \bibnamefont
  {Lykasov}}, \bibinfo {author} {\bibfnamefont {C.~J.}\ \bibnamefont
  {Pethick}}, \ and\ \bibinfo {author} {\bibfnamefont {A.}~\bibnamefont
  {Schwenk}},\ }\href {\doibase 10.1103/PhysRevC.78.045803} {\bibfield
  {journal} {\bibinfo  {journal} {Phys. Rev.}\ }\textbf {\bibinfo {volume}
  {C78}},\ \bibinfo {pages} {045803} (\bibinfo {year} {2008})},\ \Eprint
  {http://arxiv.org/abs/0808.0330} {arXiv:0808.0330 [nucl-th]} \BibitemShut
  {NoStop}%
\bibitem [{\citenamefont {Shen}\ \emph {et~al.}(2013)\citenamefont {Shen},
  \citenamefont {Gandolfi}, \citenamefont {Reddy},\ and\ \citenamefont
  {Carlson}}]{Shen:2012sa}%
  \BibitemOpen
  \bibfield  {author} {\bibinfo {author} {\bibfnamefont {G.}~\bibnamefont
  {Shen}}, \bibinfo {author} {\bibfnamefont {S.}~\bibnamefont {Gandolfi}},
  \bibinfo {author} {\bibfnamefont {S.}~\bibnamefont {Reddy}}, \ and\ \bibinfo
  {author} {\bibfnamefont {J.}~\bibnamefont {Carlson}},\ }\href {\doibase
  10.1103/PhysRevC.87.025802} {\bibfield  {journal} {\bibinfo  {journal} {Phys.
  Rev.}\ }\textbf {\bibinfo {volume} {C87}},\ \bibinfo {pages} {025802}
  (\bibinfo {year} {2013})},\ \Eprint {http://arxiv.org/abs/1205.6499}
  {arXiv:1205.6499 [nucl-th]} \BibitemShut {NoStop}%
\bibitem [{\citenamefont {Bedaque}\ and\ \citenamefont
  {Rupak}(2003)}]{Bedaque:2002xy}%
  \BibitemOpen
  \bibfield  {author} {\bibinfo {author} {\bibfnamefont {P.~F.}\ \bibnamefont
  {Bedaque}}\ and\ \bibinfo {author} {\bibfnamefont {G.}~\bibnamefont
  {Rupak}},\ }\href {\doibase 10.1103/PhysRevB.67.174513} {\bibfield  {journal}
  {\bibinfo  {journal} {Phys. Rev.}\ }\textbf {\bibinfo {volume} {B67}},\
  \bibinfo {pages} {174513} (\bibinfo {year} {2003})},\ \Eprint
  {http://arxiv.org/abs/cond-mat/0206527} {arXiv:cond-mat/0206527 [cond-mat]}
  \BibitemShut {NoStop}%
\bibitem [{\citenamefont {Rrapaj}\ \emph {et~al.}(2015)\citenamefont {Rrapaj},
  \citenamefont {Holt}, \citenamefont {Bartl}, \citenamefont {Reddy},\ and\
  \citenamefont {Schwenk}}]{Rrapaj:2014yba}%
  \BibitemOpen
  \bibfield  {author} {\bibinfo {author} {\bibfnamefont {E.}~\bibnamefont
  {Rrapaj}}, \bibinfo {author} {\bibfnamefont {J.~W.}\ \bibnamefont {Holt}},
  \bibinfo {author} {\bibfnamefont {A.}~\bibnamefont {Bartl}}, \bibinfo
  {author} {\bibfnamefont {S.}~\bibnamefont {Reddy}}, \ and\ \bibinfo {author}
  {\bibfnamefont {A.}~\bibnamefont {Schwenk}},\ }\href {\doibase
  10.1103/PhysRevC.91.035806} {\bibfield  {journal} {\bibinfo  {journal} {Phys.
  Rev.}\ }\textbf {\bibinfo {volume} {C91}},\ \bibinfo {pages} {035806}
  (\bibinfo {year} {2015})},\ \Eprint {http://arxiv.org/abs/1408.3368}
  {arXiv:1408.3368 [nucl-th]} \BibitemShut {NoStop}%
\bibitem [{\citenamefont {Kubo}(1966)}]{0034-4885-29-1-306}%
  \BibitemOpen
  \bibfield  {author} {\bibinfo {author} {\bibfnamefont {R.}~\bibnamefont
  {Kubo}},\ }\href {http://stacks.iop.org/0034-4885/29/i=1/a=306} {\bibfield
  {journal} {\bibinfo  {journal} {Reports on Progress in Physics}\ }\textbf
  {\bibinfo {volume} {29}},\ \bibinfo {pages} {255} (\bibinfo {year}
  {1966})}\BibitemShut {NoStop}%
\bibitem [{\citenamefont {Nozieres}\ and\ \citenamefont
  {Pines}(1999)}]{nozieres1999theory}%
  \BibitemOpen
  \bibfield  {author} {\bibinfo {author} {\bibfnamefont {P.}~\bibnamefont
  {Nozieres}}\ and\ \bibinfo {author} {\bibfnamefont {D.}~\bibnamefont
  {Pines}},\ }\href {https://books.google.com/books?id=q3wCwaV-gmUC} {\emph
  {\bibinfo {title} {Theory Of Quantum Liquids}}},\ Advanced Books Classics\
  (\bibinfo  {publisher} {Avalon Publishing},\ \bibinfo {year}
  {1999})\BibitemShut {NoStop}%
\bibitem [{\citenamefont {Beth}\ and\ \citenamefont
  {Uhlenbeck}(1937)}]{Beth1937915}%
  \BibitemOpen
  \bibfield  {author} {\bibinfo {author} {\bibfnamefont {E.}~\bibnamefont
  {Beth}}\ and\ \bibinfo {author} {\bibfnamefont {G.~E.}\ \bibnamefont
  {Uhlenbeck}},\ }\href {\doibase
  http://dx.doi.org/10.1016/S0031-8914(37)80189-5} {\bibfield  {journal}
  {\bibinfo  {journal} {Physica}\ }\textbf {\bibinfo {volume} {4}},\ \bibinfo
  {pages} {915 } (\bibinfo {year} {1937})}\BibitemShut {NoStop}%
\bibitem [{\citenamefont {Liu}\ \emph {et~al.}(2009)\citenamefont {Liu},
  \citenamefont {Hu},\ and\ \citenamefont {Drummond}}]{PhysRevLett.102.160401}%
  \BibitemOpen
  \bibfield  {author} {\bibinfo {author} {\bibfnamefont {X.-J.}\ \bibnamefont
  {Liu}}, \bibinfo {author} {\bibfnamefont {H.}~\bibnamefont {Hu}}, \ and\
  \bibinfo {author} {\bibfnamefont {P.~D.}\ \bibnamefont {Drummond}},\ }\href
  {\doibase 10.1103/PhysRevLett.102.160401} {\bibfield  {journal} {\bibinfo
  {journal} {Phys. Rev. Lett.}\ }\textbf {\bibinfo {volume} {102}},\ \bibinfo
  {pages} {160401} (\bibinfo {year} {2009})}\BibitemShut {NoStop}%
\bibitem [{\citenamefont {Hofmann}\ and\ \citenamefont
  {Zwerger}(2017)}]{PhysRevX.7.011022}%
  \BibitemOpen
  \bibfield  {author} {\bibinfo {author} {\bibfnamefont {J.}~\bibnamefont
  {Hofmann}}\ and\ \bibinfo {author} {\bibfnamefont {W.}~\bibnamefont
  {Zwerger}},\ }\href {\doibase 10.1103/PhysRevX.7.011022} {\bibfield
  {journal} {\bibinfo  {journal} {Phys. Rev. X}\ }\textbf {\bibinfo {volume}
  {7}},\ \bibinfo {pages} {011022} (\bibinfo {year} {2017})}\BibitemShut
  {NoStop}%
\bibitem [{\citenamefont {Horowitz}\ \emph {et~al.}(2017)\citenamefont
  {Horowitz}, \citenamefont {Caballero}, \citenamefont {Lin}, \citenamefont
  {O'Connor},\ and\ \citenamefont {Schwenk}}]{Horowitz:2016gul}%
  \BibitemOpen
  \bibfield  {author} {\bibinfo {author} {\bibfnamefont {C.~J.}\ \bibnamefont
  {Horowitz}}, \bibinfo {author} {\bibfnamefont {O.~L.}\ \bibnamefont
  {Caballero}}, \bibinfo {author} {\bibfnamefont {Z.}~\bibnamefont {Lin}},
  \bibinfo {author} {\bibfnamefont {E.}~\bibnamefont {O'Connor}}, \ and\
  \bibinfo {author} {\bibfnamefont {A.}~\bibnamefont {Schwenk}},\ }\href
  {\doibase 10.1103/PhysRevC.95.025801} {\bibfield  {journal} {\bibinfo
  {journal} {Phys. Rev.}\ }\textbf {\bibinfo {volume} {C95}},\ \bibinfo {pages}
  {025801} (\bibinfo {year} {2017})},\ \Eprint
  {http://arxiv.org/abs/1611.05140} {arXiv:1611.05140 [nucl-th]} \BibitemShut
  {NoStop}%
\end{thebibliography}%
\end{document}